\documentclass[12pt]{article}

\usepackage{a4wide,axodraw}
\usepackage{amsmath}
\usepackage{amsfonts}
\usepackage{amssymb}
\usepackage{relsize}
\usepackage{empheq}
\usepackage{cite}

\usepackage{resizegather}

\newcommand{\ep}{\varepsilon}
\def\OO{\mathcal{O}}
\def\ooS{\overline{\overline{S_1}}}
\def\oS{\overline{S_1}}
\newcommand{\Li}{\mathop{\mathrm{Li}}\nolimits}

\begin{document}

\title{
\vskip-3cm{\baselineskip14pt
\centerline{\normalsize DESY 19-139\hfill ISSN 0418-9833}
\centerline{\normalsize August 2019\hfill}}
\vskip1.5cm
Two-loop diagrams in nonrelativistic QCD with elliptics
}

\author{B.~A.~Kniehl${}^{a,}$\footnote{%
On leave of absence from II. Institut f\"ur Theoretische Physik,
Universit\"at Hamburg, Luruper Chaussee~149, 22761 Hamburg, Germany.},    
A.~V.~Kotikov${}^{b}$, A.~I.~Onishchenko${}^{b,c}$,
O.~L.~Veretin${}^{d,e}$\\
\\
${}^a$ {\normalsize\it Department of Physics, University of California at San Diego,}\\
{\normalsize\it 9500 Gilman Drive, La Jolla, CA 92093, USA}\\
${}^b$ {\normalsize\it Bogolyubov Laboratory for Theoretical Physics, JINR,}\\
{\normalsize\it 141980 Dubna (Moscow region), Russia}\\
${}^c$ {\normalsize\it Skobeltsyn Institute of Nuclear Physics, Moscow State University,}\\
{\normalsize\it 119991 Moscow, Russia}\\
${}^d$ {\normalsize\it Institut f\"ur Theoretische Physik, Universit\"at Regensburg,}\\
{\normalsize\it Universit\"atsstra\ss e 31, 93053 Regensburg, Germany}\\
${}^e$ {\normalsize\it II.~Institut f\"ur Theoretische Physik, Universit\"at Hamburg,}\\
{\normalsize\it Luruper Chaussee 149, 22761 Hamburg, Germany}}

\date{}

\maketitle

\begin{abstract}
  We consider two-loop two-, three-, and four-point diagrams
  with elliptic subgraphs involving two different masses, $m$ and $M$.
  Such diagrams generally arise in  matching procedures within nonrelativistic
  QCD and QED and are relevant, e.g., for top-quark pair production at
  threshold and parapositronium decay.
  We present the obtained results in several different representations: series
  solution with binomial coefficients, integral representation, and
  representation in terms of generalized hypergeometric functions.
  The results are valid up to terms of $\OO(\ep)$ in $d=4-2\ep$ space-time
  dimensions.
  In the limit of equal masses, $m=M$, the obtained results are written in
  terms of elliptic constants with explicit series representations.

\medskip

\end{abstract}


\newpage

\section{Introduction}
\label{sec:intro}

During the last two decades, great progress has been made in the calculation of
multiloop Feynman diagrams. This was possible thanks to the appearance of numerous new techniques. In particular, many advancements became possible due to the development of the theory of multiple polylogarithms \cite{Kummer,Nielsen,Lappo,Goncharov_MPLs_1,HarmonicPolylogs,Goncharov_MPLs_2}. A summary of their algebraic and numerical algorithms may be found in Ref.~\cite{VollingaWeinzierl}.  At special values of their arguments, polylogarithms may transform to multiple Euler--Zagier sums \cite{EulerZagier1,EulerZagier3,EulerZagier2,EulerZagier4}, ``sixth-root of unity'' constants \cite{sixroot1,Fleischer:1999mp,sixroot2,sixroot3,sixroot4,sixroot6,sixroot5,sixroot7}, multiple binomial sums \cite{sixroot3,sixroot4,Fleischer:1997bw,FleischerKotikovVeretin,Fleischer:1999hp,binomialsums1,Davydychev:2000na,binomialsums2,binomialsums4,Kotikov:2007vr,Kalmykov:2007dk,Kalmykov:2010gb,binomialsums5}, and others. Calculations yielding polylogarithms are often related to massless problems, sometimes
also to massive ones where the unitary cuts of a Feynman diagram do not cross more than two massive lines. It should be noted that, in dimensional regularization with $d=4-2\ep$ space-time dimensions, all one-loop diagrams with arbitrary masses and kinematics can, at least in principle, be expressed in terms of polylogarithms at any order in $\ep$.\footnote{%
At order $\OO(\ep)$, this was already stated in Ref.~\cite{scalar1loop}.}

Going beyond the class of polylogarithms is a challenging task, which requires additional investigations. If we are aiming at a representation of a result that
is more complicated than just a number, then we should answer the question as to
what kinds of functions are required for that.
Is it possible to classify, enumerate, and build {\it higher} functions, so that also the results at higher orders in the $\ep$ expansion, and eventually at
higher loop orders, can be expressed within that class of functions? Finally, we need a stable and fast numerical library for all such new functions introduced. The knowledge of the whole class of functions and its algebraic structure enables us to apply a very powerful method of calculation, based on differential equations \cite{diffeq1,Kotikov:1990zs,diffeq2,Kotikov:1991pm,Remiddi:1997ny}. Knowing the structure of the answer, we are able to construct an appropriate ansatz for the solutions and to solve the corresponding differential equations.
It is probably fair to say that the above program has been elaborated, to a
large extent, only for the class of polylogarithmic functions so far.
Nevertheless, there has recently been a lot of progress in understanding the
simplest functions beyond multiple polylogarithms, the so-called elliptic polylogarithms \cite{Beilinson:1994,Wildeshaus,Levin:1997,Levin:2007,Enriquez:2010,Brown:2011,Bloch:2013tra,Adams:2014vja,Bloch:2014qca,Adams:2015gva,Adams:2015ydq,Adams:2016xah,Remiddi:2017har,Broedel:2017kkb,Broedel:2017siw,Broedel:2018iwv,Broedel:2018qkq,Broedel:2019hyg,Broedel:2019tlz,Bogner:2019lfa,Broedel:2019kmn}.
However, it is already clear that, even at two loops, elliptic polylogarithms are not sufficient for all possible applications, as there can either be several elliptic curves  \cite{Adams:2018bsn,Adams:2018kez} or completely new functions present \cite{Bloch:2014qca,beyond-eMPL-1,beyond-eMPL-4,beyond-eMPL-5,beyond-eMPL-2}.

In this paper, we proceed with the study of Feynman diagrams possessing elliptic structure. We consider diagrams with only one or two independent parameters. The former case is represented by the so-called single-scale integrals, which, by dimensional reasons, can be written as a product of a scale and a numerical factor.
In the latter case, the considered diagram is expressed in terms of a function of one variable. The analytic structure of such functions can be analyzed with the help of differential equations which these functions obey. The single-scale limit is obtained by setting an existing variable to the appropriate fixed value.

To be specific, we consider diagrams for $2\to 2$ processes with
two real photons ($\gamma\gamma$) or gluons ($gg$) in the initial state and two on-shell massive fermions ($f\bar{f}$\,) in the final state. In addition, we stick to threshold kinematics. For example, we have
$\gamma(q_1) + \gamma(q_2) \to f(p_1) + \bar{f}(p_2)$, where
\begin{equation}
  q_1^2 = q_2^2 = 0, \quad p_1^2=p_2^2 \equiv p^2 =m^2, \quad s=(q_1+q_2)^2=4m^2,
\quad t=(p-q_1)^2 = -m^2 \,.
\label{kinematics}
\end{equation}
The physical motivation to study such processes with threshold kinematics is the following. Such kinematics appears, for example, in the matching procedures of QCD or QED to the corresponding nonrelativistic effective theories, such as NRQCD and NRQED, where the above processes define the corresponding hard Wilson coefficients. The NRQCD applications are related to 
heavy-quarkonium production and decays (see Refs.~\cite{quarkoniumprod5,quarkoniumprod4,quarkoniumprod3,quarkoniumprod2,quarkoniumprod1} and references cited
therein) as well as to near-threshold $t\bar{t}$ production (see Refs.~\cite{Adams:2018bsn,Adams:2018kez,topprod6,topprod11,topprod5,topprod14,topprod10,topprod9,topprod4,topprod13,topprod8,topprod7,topprod12,topprod3,topprod2,topprod1} and references cited therein), which offers the opportunity to considerably improve the accuracy of $t$-quark mass measurements at the CERN Large Hadron Collider and future linear colliders. As for NRQED applications, the best known one is probably the calculation of the parapositronium decay rate \cite{Kniehl:2000dh,parapositronium1,parapositronium3,parapositronium2,Kniehl:2008dt}.

%
%
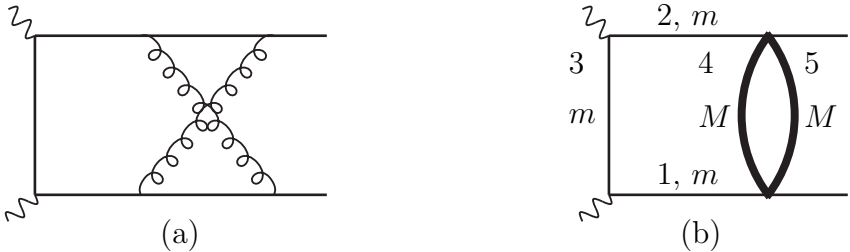
\begin {figure} [htbp]
~~~~~~~
\begin{picture}(150,100)(0,0)
\SetWidth{1.0}
\Line(20,20)(130,20)
\Line(20,80)(130,80)
\Line(20,20)(20,80)
\SetWidth{0.7}
\Gluon(60,20)(110,80){3}{8}
\Gluon(60,80)(110,20){3}{8}
\Photon(20,20)(10,10){3}{2}
\Photon(20,80)(10,90){3}{2}
\Text(75,5)[c]{(a)}
\end{picture}
~~~~~~~~~~
\begin{picture}(150,100)(0,0)
\SetWidth{1.0}
\Line(40,20)(130,20)
\Line(40,80)(130,80)
\Line(40,20)(40,80)
\SetWidth{3.0}
\CArc(140,50)(50,142,218)
\CArc(60,50)(50,-38,38)
\SetWidth{0.7}
\Photon(40,20)(30,10){3}{2}
\Photon(40,80)(30,90){3}{2}
\Text(70,27)[c]{$1,\,m$}
\Text(70,87)[c]{$2,\,m$}
\Text(25,70)[l]{$3$}
\Text(25,50)[l]{$m$}
\Text(74,70)[l]{$4$}
\Text(74,50)[l]{$M$}
\Text(114,70)[l]{$5$}
\Text(114,50)[l]{$M$}
\Text(75,5)[c]{(b)}
\end{picture}
\label{nonplanar}
\caption{
(a) Two-loop nonplanar diagram contributing to NNLO matching in nonrelativistic field theory and (b) diagram obtained from it by contracting the massless-propagator lines. An additional mass $M$ is introduced to simplify the actual calculations (see discussion in the text).
}
\end{figure}

As already mentioned above, there are no new functions beyond polylogarithms at
the one-loop level. So, the elliptic structures first appear at next-to-next-to-leading order (NNLO).
One of the most complicated diagrams containing elliptic structure
is the nonplanar one shown in Fig.~\ref{nonplanar}(a). Applying integration-by-parts (IBP) identities \cite{ibp,Chetyrkin:1981qh}, this diagram can be reduced to a sum of so-called master integrals with rational coefficients. When both massless lines are contracted, we obtain the diagram shown in Fig.~\ref{nonplanar}(b) upon the replacement $M=m$. In this paper, we evaluate the latter diagram and all its subtopologies obtained by contracting some of its lines. Moreover, as a by product, we even solve the more general problem with two different masses, as shown
in Fig.~\ref{nonplanar}(b). 

Here and in the following, we work in dimensional regularization with
$d=4-2\ep$ space-time dimensions.
In our previous paper \cite{sunsetsNRQCD}, we presented the results for sunset diagrams obtained by contracting lines 1 and 2 in Fig.~\ref{nonplanar}(b).
The results of Ref.~\cite{sunsetsNRQCD} were obtained up to and including the $\OO(\ep)$ terms and represented as fast converging series in the mass ratio $x=m^2/M^2$. The $\OO(1)$ terms may also be found in Section~\ref{seriesRep}. It is then easy to rewrite a given such series in terms of generalized hypergeometric functions ${}_{p}F_{p-1}$ \cite{Abramowitz}, as
\begin{eqnarray}
\lefteqn{e^{2\ep\gamma_{E}}\left(M^{2}\right)^{2\ep+1}J_{01022}^{mM}=  _{4}F_{3}\left(\begin{array}{c}
\frac{1}{2},1,1,1\\
\frac{3}{4},\frac{5}{4},\frac{3}{2}
\end{array}\Bigg|-\frac{x^{2}}{4}\right)}\nonumber \\
&&{} +\frac{1}{4}\,\frac{d}{d\delta}\left[ \frac{x^{1-2\delta}(1-2\delta)^{2}}{(1-\delta)(3-16\delta)}{}_{4}F_{3}\left(\begin{array}{c}
1,1-\delta,\frac{3}{2}-\delta,\frac{3}{2}-\delta\\
\frac{5}{4}-\delta,\frac{7}{4}-\delta,2-\delta
\end{array}\Bigg|-\frac{x^{2}}{4}\right)\right]_{\delta=0}+\OO(\ep)\,,
\nonumber\\
\lefteqn{e^{2\ep\gamma_{E}}\left(M^{2}\right)^{2\ep}J_{01012}^{mM}=  -\frac{1}{2\ep^{2}}-\frac{1}{2\ep}-\frac{1}{2}(1+\zeta_{2})+\frac{x}{4}(5-2\ln x)}\nonumber\\
&&{}+\frac{x^{2}}{9}{}_{4}F_{3}\left(\begin{array}{c}
1,1,1,\frac{3}{2}\nonumber\\
\frac{5}{4},\frac{7}{4},\frac{5}{2}
\end{array}\Bigg|-\frac{x^{2}}{4}\right)\\
&&{}-\frac{d}{d\delta}\left[ \frac{x^{3-\delta}(1-\delta)^{2}}{(4-\delta)(15-46\delta)}{}_{4}F_{3}\left(\begin{array}{c}
1,\frac{3}{2}-\frac{\delta}{2},\frac{3}{2}-\frac{\delta}{2},2-\frac{\delta}{2}\\
\frac{7}{4}-\frac{\delta}{2},\frac{9}{4}-\frac{\delta}{2},3-\frac{\delta}{2}
\end{array}\Bigg|-\frac{x^{2}}{4}\right)\right]_{\delta=0}+\OO(\ep)\,,
\nonumber\\
\lefteqn{e^{2\ep\gamma_{E}}\left(M^{2}\right)^{2\ep-1}J_{01011}^{mM}=  -\frac{1}{\ep^{2}}(1+\frac{x}{2})+\frac{1}{\ep}\left[-3+\frac{1}{4}(-7+4\ln x)\right]-7-\zeta_{2}}\nonumber\\
&&{}+\frac{x}{8}\left(5-4\zeta_{2}+8\ln x-4\ln^{2}x\right)-\frac{2x^{2}}{9}{}_{4}F_{3}\left(\begin{array}{c}
\frac{1}{2},1,1,1\\
\frac{5}{4},\frac{7}{4},\frac{5}{2}
\end{array}\Bigg|-\frac{x^{2}}{4}\right)\nonumber\\
&&{}-2\frac{d}{d\delta}\left[ \frac{x^{3+\delta}(1+\delta)^{2}}{(2+\delta)(4+\delta)(15+46\delta)}{}_{4}F_{3}\left(\begin{array}{c}
1,1+\frac{\delta}{2},\frac{3}{2}+\frac{\delta}{2},\frac{3}{2}+\frac{\delta}{2}\\
\frac{7}{4}+\frac{\delta}{2},\frac{9}{4}+\frac{\delta}{2},3+\frac{\delta}{2}
\end{array}\Bigg|-\frac{x^{2}}{4}\right)\right]_{\delta=0}+\OO(\ep)\,,\
\label{eq:gehyfu}
\end{eqnarray}
where $\zeta_n=\zeta(n)$ denotes Riemann's zeta function. 
A similar form was found for the Baxter $Q$ function in Refs.~\cite{BaxterMellin,Beccaria:2009rw}. Following our study in Ref.~\cite{sunsetsNRQCD}, it was shown in Ref.~\cite{KalmykovKniehl} that a similar hypergeometric representation is applicable for the considered sunset diagrams even without $\ep$ expansion. In the present paper, we extend the results of Ref.~\cite{sunsetsNRQCD} for the case of three- and four-point diagrams (see Fig.~\ref{nonplanar}) with $\OO(1)$ accuracy.

This paper is organized as follows.
In Section~\ref{relation}, we recall the central relation between the one- and two-loop diagrams, Eq.~(\ref{master}). Section~\ref{1loop} contains the details of the one-loop calculations. In Section~\ref{2loop}, we derive the series, integral, and hypergeometric representations of the considered two-loop diagrams in the general case of non-equal masses, $M\neq m$.
Subsection~\ref{equalmass} contains our results for the equal-mass case. Finally, in Section~\ref{conclusion}, we present our conclusions.
In Appendix~\ref{diffeqFrobenius}, we explain the Frobenius solution for a system of differential equations.

\section{Relation between one- and two-loop diagrams}
\label{relation}

As explained in Section~\ref{sec:intro}, all two-loop Feynman diagrams considered in this paper have self-energy insertions built from two massive propagators
as indicated in Fig.~\ref{nonplanar}(b). These can be reduced to one-loop diagrams by representing the loop with the two massive propagators as an integral whose integrand contains a new propagator with a mass that depends on the variable of integration \cite{Fleischer:1997bw,FleischerKotikovVeretin,diffeq1,Kotikov:1990zs,sunsetsNRQCD,Kniehl:2005yc}.
Graphically, this procedure has the following form:
\begin{equation}
\mbox{{
\begin{picture}(60,30)(0,13)
\SetWidth{1.0}
\SetWidth{2.0}
\Curve{(5,15)(30,25)(55,15)}
\Curve{(5,15)(30,5)(55,15)}
\SetWidth{1.0}
\Line(5,15)(-5,15)
\Line(55,15)(65,15)
\Text(30,27)[b]{$\scriptstyle a,\,M^2_1$}
\Text(30,3)[t]{$\scriptstyle b,\,M^2_2$}
\Text(0,12)[t]{$q$}
\end{picture}
}}
\hspace{2mm} ~=~
i^{1+d} \frac{\Gamma(a+b-d/2)}{\Gamma(a)\,\Gamma(b)}
   \int\limits_0^1
\frac{{\rm d}s}{s^{a+1-d/2}(1-s)^{b+1-d/2}}
\times\hspace{3mm}
\raisebox{1mm}{{
\begin{picture}(70,30)(0,4)
\SetWidth{2.0}
\Line(5,5)(65,5)
\Vertex(5,5){2}
\SetWidth{1.0}
\Vertex(65,5){2}
\Line(5,5)(-5,5)
\Line(65,5)(75,5)
\Text(33,7)[b]{$\scriptstyle a+b-d/2$}
\Text(33,7)[b]{}
\Text(33,1)[t]{$\scriptstyle M^2_1/s+M^2_2/(1-s)$}
\Text(-3,-5)[b]{$q$}
\end{picture}
}}
\hspace{3mm},
\label{master}
\end{equation}
where the loop involving the two propagators with mass square $M^2$ is replaced by one propagator with mass square $M^2_1/s+M_2^2/(1-s)$.
The numbers $a$ and $b$, indicating the powers of the scalar propagators, are called the propagator indices.

Equation~(\ref{master}) is easily derived from the Feynman parameter
representation and was introduced in Euclidean and Minkowski spaces in
Refs.~\cite{diffeq1,Kotikov:1990zs} and \cite{Fleischer:1997bw,FleischerKotikovVeretin}, respectively.
Here, we work in Minkowski space and thus follow Refs.~\cite{Fleischer:1997bw,FleischerKotikovVeretin}.

As in Ref.~\cite{sunsetsNRQCD}, we adopt the following strategy. First, applying Eq.~(\ref{master}), we write expressions for the considered two-loop diagrams as integrals of one-loop diagrams involving propagators with masses that depend on the variable of integration. Next, in Section~\ref{1loop}, we evaluate the one-loop integrals thus obtained and then, in Section~\ref{2loop}, use these expressions to reconstruct the results for the two-loop diagrams involving two different masses with the help of Eq.~(\ref{master}). A similar strategy was also adopted for the calculation of certain four-loop tadpole diagrams in Ref.~\cite{Kniehl:2005yc}.

\section{One-loop integrals}
\label{1loop}

\begin {figure} [htbp]
\centerline{
\hspace*{-10mm}
\begin{picture}(110,80)(0,0)
\SetWidth{1.0}
\Line(70,20)(100,20)
\Line(70,80)(100,80)
\SetWidth{1.0}
\CArc(110,50)(50,142,218)
\SetWidth{3.0}
\CArc(30,50)(50,-38,38)
\SetWidth{0.7}
\Photon(50,85)(70,80){3}{3}
\Photon(50,10)(70,20){3}{3}
\Text(75,5)[c]{(a)}
\end{picture}
\begin{picture}(150,100)(0,0)
\SetWidth{1.0}
\Line(40,20)(130,20)
\Line(40,20)(100,80)
\Line(100,80)(130,80)
\SetWidth{4.0}
\Line(100,20)(100,80)
\SetWidth{0.7}
\Photon(40,20)(30,10){3}{2}
\Photon(100,80)(80,85){3}{2}
\Text(75,5)[c]{(b)}
\end{picture}
\begin{picture}(150,100)(0,0)
\SetWidth{1.0}
\Line(40,20)(130,20)
\Line(40,80)(130,80)
\Line(40,20)(40,80)
\SetWidth{4.0}
\Line(110,20)(110,80)
\SetWidth{2.0}
\SetWidth{0.7}
\Photon(40,20)(30,10){3}{2}
\Photon(40,80)(30,90){3}{2}
\Text(80,87)[c]{$1,\,m$}
\Text(80,27)[c]{$2,\,m$}
\Text(25,60)[l]{$3$}
\Text(25,45)[l]{$m$}
\Text(114,60)[l]{$4$}
\Text(114,45)[l]{$M$}
\Text(75,5)[c]{(c)}
\end{picture}
}
\caption{Typical one-loop diagrams resulting from the IBP reduction of the
integral in Eq.~(\ref{Iabcd}). The thin and thick solid lines represent scalar propagators with masses $m$ and $M$, respectively. In the case of the box diagram (c), also the line numbering is indicated.}
\label{oneloop}
\end{figure}
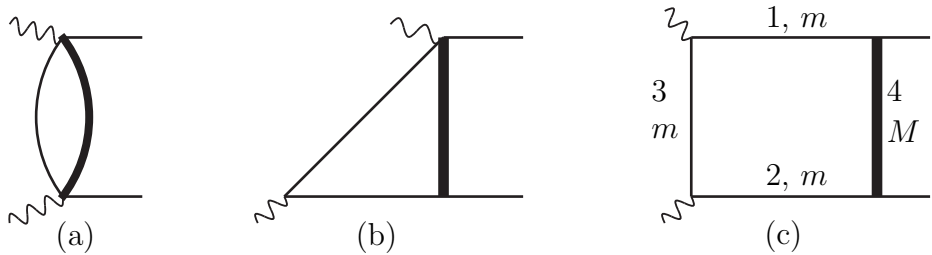

Let us consider the following one-loop box integral with two different masses, $m$ and $M$, and propagator indices $a_1,a_2,a_3,a_4$:
\begin{eqnarray}
  I_{a_1a_2a_3a_4}^{mmmM} &=&I_{a_1a_2a_3a_4}^{mM} \nonumber\\
  &=&\int\frac{d^d l}{\pi^{d/2}}
    \frac{1}{[(l+q_1)^2-m^2]^{a_1}[(l-q_2)^2-m^2]^{a_2}[l^2-m^2]^{a_3}[(l+q_1-p)^2-M^2]^{a_4}}\,,\quad
\label{Iabcd}
\end{eqnarray}
where the kinematics in Eq.~(\ref{kinematics}) is implied. Using IBP, the integral in Eq.~(\ref{Iabcd}) can always be reduced to a set of scalar master integrals with propagator indices 1 or 0. Some of these master integrals are shown in Fig.~\ref{oneloop}.

Following the line numbering in Fig.~\ref{oneloop}(c), the diagrams shown in Figs.~\ref{oneloop}(a)--(c) correspond to the integrals $I^{mmmM}_{0011}$, $I^{mmmM}_{0111}$, and $I^{mmmM}_{1111}$, respectively. Notice that, in the course of the reduction, more master integrals
appear: two bubbles, $I_{1000}$ and $I_{0001}$, two more vertices, $I_{1110}$ and $I_{1101}$, and one more self-energy, $I_{0110}$. However, all these additional integrals only contain simple polylogarithmic
structures, and their two-loop counterparts have no elliptic content. We omit these diagrams in our discussion,
except for those that enter as subgraphs in the integrals $I^{mmmM}_{0111}$ and $I^{mmmM}_{1111}$.

\subsection{Two-point case}

In the case of the two-point diagrams, IBP provides us with the relations
\begin{eqnarray}
I_{0101}^{mM} (d-3) \left(M^2 -2m^2\right) &=& M^2 I_{0002}^{M}
-2m^2 I_{0020}^{m} - M^2 \left(M^2-4m^2\right) I_{0102}^{mM} \, ,
\nonumber\\
I_{0011}^{mM}  (d-3) M^2 &=& \left(M^2 +2m^2\right) I_{0002}^{M}
-2m^2 I_{0020}^{m} - (4m^4+M^4)  I_{0012}^{mM} \, ,
\label{2pointIBP.2}
\end{eqnarray}
which can be considered as the differential equations for the initial diagrams
$ I_{0101}^{mM}$ and $I_{0011}^{mM}$, as
\begin{equation}
I_{0102}^{mM} = \frac{d}{dM^2} I_{0101}^{mM},\qquad
I_{0012}^{mM} = \frac{d}{dM^2} I_{0011}^{mM} \, .
\end{equation}
The integrals $I_{000a}^{M}$ and $I_{00a0}^{m}$ in Eq.~\eqref{2pointIBP.2} are tadpoles with masses $M$ and $m$, respectively,
\begin{equation}
I_{000a}^{M} = \frac{i (-1)^a}{(M^2)^{a-d/2}} \,
\frac{\Gamma(a-d/2)}{\Gamma(a)},\qquad I_{00a0}^{m} = \frac{i (-1)^a}{(m^2)^{a-d/2}} \,
\frac{\Gamma(a-d/2)}{\Gamma(a)}
\, .
\end{equation}
Similarly, using IBP and substituting $m=1$ and ${d}/{dM^{2}}=-x^{2}{d}/{dx}$, we can deduce a differential equation for $I_{12}^{(1)}=I_{0102}^{mM}$,
\begin{equation}
(1-2x)(1-4x)\frac{d}{dx}I_{12}^{(1)}-4xI_{12}^{(1)}+i(1-2x+2x\ln x)=0\, ,
\label{eq:i12}
\end{equation}
which is valid up to $\OO(\ep)$ corrections. Equation~\eqref{eq:i12} is easy to solve and, taking the boundary conditions at $x=0$ into account, we have
\begin{equation}
\frac{M^{2}}{i}I_{12}^{(1)}=  \frac{\ln x}{2x}+\frac{(1-2x)}{2x\sqrt{1-4x}}\ln\frac{1+\sqrt{1-4x}}{1-\sqrt{1-4x}}+\OO(\ep)\,,
\end{equation}
or, in terms of a series solution, 
\begin{equation}
\frac{M^{2}}{i}I_{12}^{(1)}=  -1-\sum_{n=1}^{\infty}\left(\begin{array}{c}
2n\\
n
\end{array}\right)\frac{n}{n+1}x^{n}\left[ \ln x-2S_{1}(n-1)+2S_{1}(2n-1)-\frac{1}{n+1}\right] +\OO(\ep)\,.
\end{equation}
Now, IBP relations allow us to write down answers for the $I_{a_1a_2}^{(1)}$ integrals with other values of propagator indices. For example, we have 
\begin{eqnarray}
  \frac{M^{2}}{i}I_{21}^{(1)}&= & -\frac{\ln x}{2x}-\frac{1}{2x\sqrt{1-4x}}\ln\frac{1+\sqrt{1-4x}}{1-\sqrt{1-4x}}+\OO(\ep)\,,
\nonumber\\  
\frac{M^{2}}{i}I_{21}^{(1)}&=& 1 +\ln x
+ \sum_{n=1}^{\infty} \binom{2n}{n}\frac{1+2n}{1+n}x^n
\nonumber\\
&&{}\times\left[
\ln x + 2 S_1 (2n-1) - 2 S_1 (n-1) + \frac{2}{2n+1} - \frac{1}{n+1} - \frac{1}{n}
\right] + \OO (\ep)\,,
\nonumber\\
e^{\ep\gamma_{E}}\frac{(M^{2})^{\ep}}{i}I_{11}^{(1)}&= & \frac{1}{\ep}+2+\frac{1-2x}{2x}\ln x+\frac{\sqrt{1-4x}}{2x}\ln\frac{1+\sqrt{1-4x}}{1-\sqrt{1-4x}}+\OO(\ep)\,,
\nonumber\\
e^{\ep\gamma_{E}}\frac{(M^{2})^{\ep}}{i}I_{11}^{(1)} &=& \frac{1}{\ep} + 1
+ \sum_{n=1}^{\infty}\binom{2n}{n}\frac{x^n}{1+n}
\nonumber\\ 
&&{}\times\left[
\ln x + 2 S_1 (2n-1) - 2 S_1 (n-1) - \frac{1}{n} - \frac{1}{n+1}
\right] +\OO(\ep)\,.
\end{eqnarray}
For $m=M$ ($x=1$), we have 
\begin{eqnarray}
\frac{M^{2}}{i}I_{12}^{(1)} &=&  \frac{M^{2}}{i}I_{21}^{(1)} = -\frac{\pi}{3\sqrt{3}} + \OO(\ep)\, , \nonumber\\
e^{\ep\gamma_{E}}\frac{(M^{2})^{\ep}}{i}I_{11}^{(1)}&=&  \frac{1}{\ep} + 2 - \frac{\pi}{\sqrt{3}} + \OO(\ep)\, .
\end{eqnarray}
In the case of the integral $I_{12}^{(2)}=I_{0012}^{mM}$, the application of IBP relations leads to the following differential equation:
\begin{equation}
\left(1+4x^{2}\right)\frac{d}{dx}I_{12}^{(2)}+4xI_{12}^{(2)}+i(1+2x+2x\ln x)=0\,,
\label{eq:i12_2}
\end{equation}
which is again valid up to $\OO(\ep)$ terms. The solution of Eq.~\eqref{eq:i12_2}
is straightforward and, taking the boundary conditions at $x=0$ into account,
yields
\begin{equation}
  \frac{M^{2}}{i}I_{12}^{(2)} = -\frac{1}{2x}\ln x
+ \frac{1}{4 x \sqrt{1+4x^2}}
\left(
\ln\frac{\sqrt{1+4x^2}-1}{\sqrt{1+4x^2}+1}
 + \ln\frac{\sqrt{1+4x^2}-2x}{\sqrt{1+4x^2}+2x}
\right) + \OO(\ep)\,,
\end{equation}
or, in terms of a series solution,
\begin{eqnarray}
\frac{M^{2}}{i}I_{12}^{(2)}&=&  -1-\sum_{n=1}^{\infty}\frac{(-16x^{2})^{n}}{(2n+1)\left(\begin{array}{c}
	2n\\
	n
  \end{array}\right)}\nonumber \\
&&{}+\frac{1}{2x}\sum_{n=1}^{\infty}\left(\begin{array}{c}
2n\\
n
\end{array}\right)(-x^{2})^{n}\left[ \ln x+S_{1}(2n-1)-S_{1}(n-1)-\frac{1}{2n}\right] + \OO(\ep)\,.\quad
\end{eqnarray}
The expressions for the $I^{(2)}_{a_1 a_2}$ integrals with other values of propagator indices can again be obtained with the help of IBP relations. For example, for $I_{21}^{(2)}$ and $I_{11}^{(2)}$, we have
\begin{eqnarray}
  \frac{M^{2}}{i}I_{21}^{(2)}&= & (2x-1)\frac{M^{2}}{i}I_{12}^{(2)}+\ln x+\OO(\ep)\,,
  \nonumber\\
e^{\ep\gamma_{E}}\frac{(M^{2})^{\ep}}{i}I_{11}^{(2)}&= & \frac{1}{\ep}+\left(1+4x^{2}\right)\frac{M^{2}}{i}I_{12}^{(2)}+2(1+x\ln x)+\OO(\ep)\,.
\end{eqnarray}
In the limit $m=M$ ($x=1$), we have 
\begin{eqnarray}
  \frac{M^{2}}{i}I_{12}^{(2)} &=&  \frac{M^{2}}{i}I_{21}^{(2)} = -\frac{2}{\sqrt{5}}\ln\frac{1+\sqrt{5}}{2} + \OO(\ep)\, ,
  \nonumber\\
e^{\ep\gamma_{E}}\frac{(M^{2})^{\ep}}{i}I_{11}^{(2)}&= & \frac{1}{\ep} + 2 - 2\sqrt{5}\ln\frac{1+\sqrt{5}}{2} + \OO(\ep)\, .
\end{eqnarray}

\subsection{Three-point case}

In the case of three-point diagrams, IBP yields the relation
\begin{equation}
I_{011a}^{mM} = \frac{1}{2m^2 (a-1)} \left[
(a-1) I_{001a} + I_{002a-1} - (a-1) I_{010a} - I_{020a-1}
\right]\, ,
\end{equation}
which, for example, allows us to obtain expressions for the integral
$I_{112}^{(3)} = I_{0112}^{mM}$,
\begin{eqnarray}
\frac{(M^2)^2}{i} I_{112}^{(3)} &=& \frac{1}{2 x \sqrt{1-4x}}\ln
\frac{1+\sqrt{1-4x}}{1-\sqrt{1-4x}} 
+ \frac{1}{4 x \sqrt{1+4x^2}}
\nonumber\\
&&{}\times\left(
\ln\frac{\sqrt{1+4x^2}-1}{\sqrt{1+4x^2}+1}
 + \ln\frac{\sqrt{1+4x^2}-2x}{\sqrt{1+4x^2}+2x}
\right) + \OO (\ep)\,,
\end{eqnarray}
or, in terms of a series solution,
\begin{eqnarray}
\frac{(M^2)^2}{i} I_{112}^{(3)} &=& -1  -\frac{1}{2x}\sum_{n=1}^{\infty}\binom{2n}{n}x^n \left[
\ln x + 2 S_1 (2n-1) - 2 S_1 (n-1) - \frac{1}{n}
\right] \nonumber\\
&&{}- \sum_{n=1}^{\infty}\frac{(-16x^{2})^{n}}{(2n+1)\left(\begin{array}{c}
2n\\ n
  \end{array}\right)} + \frac{1}{2x}\sum_{n=1}^{\infty} \binom{2n}{n} \left(-x^2\right)^n
\nonumber\\
&&{}\times
\left[
\ln x + S_1 (2n-1) - S_1 (n-1) - \frac{1}{2n}
\right] + \OO (\ep)\,.
\end{eqnarray}
To obtain an expression for the $I_{111}^{(3)}$ integral, we may consider the following, easy-to-obtain differential equation:\footnote{As before, this follows from IBP relations and is valid up to $\OO (\ep)$ terms.}
\begin{equation}
\frac{d}{dx}I_{111}^{(3)}+\frac{1}{x(1-2x)}I_{12}^{(1)}+\frac{1}{x}I_{12}^{(2)}-\frac{i\ln x}{1-2x}=0\,.
\label{eq:i111_3}
\end{equation}
The solution of Eq.~\eqref{eq:i111_3}, with account of the corresponding boundary conditions at $x=0$,\footnote{In the limit $M^2\to\infty$, the integral $I_{111}^{(3)}$ goes to zero.} is given by
\begin{equation}
\frac{M^2}{i} I_{111}^{(3)} = \frac{1}{4x}\left[ \ln^2
\frac{1+\sqrt{1-4x}}{1-\sqrt{1-4x}}
-\frac{1}{4}\ln^2
\frac{\sqrt{1+4x^2}-1}{\sqrt{1+4x^2}+1}
- 2I_{1}(x)
\right] + \OO(\ep)\, ,
\end{equation}
where
\begin{equation}
  I_{1}(x)=  \textrm{Li}_{2}(-R(x))-\Li_{2}(R(x))\,,\qquad
  R(x) = \frac{1-\sqrt{S(x)}}{1+\sqrt{S(x)}}\, ,\qquad S(x) = \frac{\sqrt{1+4x^2}-2x}{\sqrt{1+4x^2}+2x}\, .
\label{I1x}
\end{equation}
In terms of a series, the solution takes form
\begin{eqnarray}
\frac{M^2}{i} I_{111}^{(3)} &=& 1 + \ln x + \sum_{n=1}^{\infty}\frac{(-16x^{2})^{n}}{(1+2n)^2\left(\begin{array}{c}
	2n\\ n
  \end{array}\right)}
+ \sum_{n=1}^{\infty}\binom{2n}{n}\frac{(1+2n)}{(1+n)^2}x^n
\nonumber\\
&&{}\times\left[
\ln x + 2 S_1 (2n-1) - 2 S_1 (n-1) + \frac{2}{2n+1} - \frac{2}{n+1} - \frac{1}{n}
\right] \nonumber\\
&&{}+ \frac{1}{4x}\sum_{n=1}^{\infty}\binom{2n}{n}\frac{\left(-x^2\right)^n}{n}\left[
\frac{1}{n} + S_1 (n-1) - S_1 (2n-1) -\ln x
\right] + \OO (\ep)\, .
\end{eqnarray}
In the limit $m=M$ ($x=1$), we have 
\begin{eqnarray}
\frac{M^2}{i}I_{111}^{(3)} &=& \frac{1}{2}\Li_2 \left(\frac{\sqrt{5}-1}{2}\right)
- \frac{1}{2}\Li_2 \left(\frac{1-\sqrt{5}}{2}\right) - \frac{1}{16}\ln^2
\frac{\sqrt{5}+1}{\sqrt{5}-1}  - \frac{\pi^2}{9}  + \OO (\ep)\,, \nonumber\\
\frac{(M^2)^2}{i}I_{112}^{(3)} &=& \frac{\pi}{3\sqrt{3}} - \frac{2}{\sqrt{5}}\ln
\frac{1+\sqrt{5}}{2}
+ \OO (\ep) \, .
\end{eqnarray}

\subsection{Four-point case}

The solution of the four-point case is simple. In this case, we may exploit the
fact that the propagators of the integral $I_{a_1 a_2 a_3 a_4}^{m M}$ in Eq.~\eqref{Iabcd} are linearly dependent in the special kinematics of Eq.~\eqref{kinematics} and so that diagrams with four propagators are reduced to ones with at least one propagator less. Namely, we have the relation
\begin{equation}
1 = \frac{(l+q_1)^2 + (l-q_2)^2 - 2 (l+q_1-p)^2 + 2 (M^2-m^2)}{2 M^2}\, ,
\end{equation}
which can be used to obtain the following expressions:
\begin{eqnarray}
I_{1111}^{mM} &=& \frac{1}{M^2} \left( I_{0111}^{mM}-I_{1110}^{mM} \right)
= \frac{1}{M^2} \left( I_{111}^{(3)}-I_{1110}^{mM} \right)
\, , \nonumber\\ 
I_{1112}^{mM} &=& \frac{1}{M^2} \left(
I_{0112}^{mM} - \frac{1}{M^2} I_{0111}^{mM} + \frac{1}{M^2} I_{1110}^{mM}
\right) = \frac{1}{M^2} \left(
I_{112}^{(3)} - \frac{1}{M^2} I_{111}^{(3)} + \frac{1}{M^2} I_{1110}^{mM}
\right) \, ,
\end{eqnarray} 
where
\begin{equation}
\frac{m^2}{i}I_{1110}^{mM} = -\frac{\pi^2}{8} + \OO (\ep)\, .
\end{equation}

\section{Two-loop diagrams}
\label{2loop}

We are now ready to evaluate the two-loop diagrams $J_{010a_1a_2}^{mM}$,
$J_{011a_1a_2}^{mM}$, and $J_{111a_1a_2}^{mM}$, which can be obtained from $J_{b_1b_2b_3a_1a_2}^{mM}$ shown in Fig. \ref{nonplanar}(b). The latter integral has the expression
\begin{eqnarray}
J_{b_1b_2b_3a_1a_2}^{mM} &=& \int \frac{{\rm d}^dk\,{\rm d}^dk_1}{\pi^d}\,
  \frac{1}{(k^2-m^2)^{b_3}[(k+q_1)^2-m^2]^{b_1}[(k-q_2)^2-m^2]^{b_2}}  \nonumber \\
&&{}\times  \frac{1}{[(k_1+q_1/2-q_2/2)^2-M^2]^{a_1}
[(k_1-k)^2+M^2]^{a_2}} \, ,
\label{2loopA4}
\end{eqnarray}
where again the special kinematics of Eq.~\eqref{kinematics} is implied. Exploiting the rule of Eq.~\eqref{master}, we replace the $k_1$ integration in
Eq.~\eqref{2loopA4} with the integration over the variable $s$ from the
corresponding one-loop integral $I_{...a}^{mM}$ with the replacement $M^2 \to \mu^2=M^2/[s(1-s)]$.
Indeed, calling the above two-loop integral $J_{\dots a_1a_2}^{mM}$
and the corresponding one-loop diagram $I_{\dots a}^{mM}$,
we have
\begin{equation}
J_{\dots a_1a_2}^{mM} = \frac{\Gamma(a)}{\Gamma(a_1)\Gamma(a_2)} \,
\int^1_0 ds s^{\tilde{a}_1-1} (1-s)^{\tilde{a}_2-1}
I_{\dots a}^{m\mu}\, ,
\label{2loopA1}
\end{equation}
where $\tilde{a}_i=d/2-a_i$ and $a=a_1+a_2-d/2$. We note that, strictly speaking, in Eq.~\eqref{2loopA1}, we should use the one-loop integrals $I_{\dots, k+\ep}^{mM}$, where $k=0,1,2$. On the other hand, in Section~\ref{1loop}, we presented results for the $I_{\dots k}^{mM}$ integrals with $\OO (1)$ accuracy only. However, the latter are sufficient to determine the most complicated contributions from the two-loop diagrams, containing series in $n$. The less complicated terms can be obtained directly from the two-loop diagrams, either by using asymptotic expansions or differential equations. Moreover, the knowledge of the Frobenius solutions of the differential equations together with the ans\"atze for the occurring sums is sufficient to determine the required expressions for the two-loop diagrams (see Appendix~\ref{diffeqFrobenius}).

\subsection{Series representations}
\label{seriesRep}

Let us start with the two-point sunset diagrams. It is known, e.g.\ from Ref.~\cite{Tarasov:1997kx}, that for the special kinematics in Eq.~\eqref{kinematics}
there exist three master integrals. Choosing the latter to be $J_{01011}^{mM}$, $J_{01012}^{mM}$, and
$J_{01022}^{mM}$ and following the above general procedure, we obtain after some calculations
\begin{eqnarray}
e^{2\ep\gamma_{E}}\left(M^{2}\right)^{2\ep+1}J_{01022}^{mM}&= & J_{122}^{(1)}+J_{122}^{(2)} + \OO(\ep)\,,\nonumber\\
e^{2\ep\gamma_{E}}\left(M^{2}\right)^{2\ep}J_{01012}^{mM}&= & -\frac{1}{2}\left(\frac{1}{\ep^{2}}+\frac{1}{\ep}+1+\zeta_{2}\right)+J_{112}^{(1)}+J_{112}^{(2)} + \OO(\ep)\,,\nonumber\\
e^{2\ep\gamma_{E}}\left(M^{2}\right)^{2\ep-1}J_{01011}^{mM}&= & -\left(\frac{1}{\ep^{2}}+\frac{3}{\ep}+7+\zeta_{2}\right)\nonumber \\
&&{} +x\left[-\frac{1}{2}\left(\frac{1}{\ep^{2}}+\frac{7}{2\ep}-\frac{5}{4}+\zeta_{2}\right)+\left(\frac{1}{\ep}+1\right)\ln x-\frac{1}{2}\ln^{2}x\right] \nonumber \\ &&{} +J_{111}^{(1)}+J_{111}^{(2)} + \OO(\ep)\,,
\end{eqnarray}
where 
\begin{eqnarray}
J_{122}^{(1)}&= & \frac{1}{2x}\sum_{n=1}^{\infty}\frac{\left(\begin{array}{c}
	2n\\
	n
	\end{array}\right)}{\left(\begin{array}{c}
	4n\\
	2n
  \end{array}\right)}\,\frac{\left(-x^{2}\right)^{n}}{n}\left( \frac{1}{2n}+S_1-3\oS+2\ooS-\ln x\right)\,,
\nonumber\\
J_{122}^{(2)}&= & \frac{1}{4x^{2}}\sum_{n=1}^{\infty}\frac{\left(-16x^{2}\right)^{n}}{\left(\begin{array}{c}
	2n\\
	n
	\end{array}\right)\left(\begin{array}{c}
	4n\\
	2n
  \end{array}\right)}\,\frac{1-4n}{n^{2}(2n-1)}\,,
\nonumber\\
J_{112}^{(1)}&= & \frac{x}{4}(5-2\ln x) \nonumber \\ &&{}+\frac{x}{2}\sum_{n=1}^{\infty}\frac{\left(\begin{array}{c}
	2n\\
	n
	\end{array}\right)}{\left(\begin{array}{c}
	4n\\
	2n
  \end{array}\right)}\,\frac{\left(-x^{2}\right)^{n}}{(n+1)(4n+1)}\left( \frac{1}{2(n+1)}+\frac{2}{1+4n}+S_{1}-3\oS+2\ooS-\ln x\right)\,,
\nonumber\\
J_{112}^{(2)}&= & -\frac{1}{4}\sum_{n=1}^{\infty}\frac{\left(-16x^{2}\right)^{n}}{\left(\begin{array}{c}
	2n\\
	n
	\end{array}\right)\left(\begin{array}{c}
	4n\\
	2n
  \end{array}\right)}\,\frac{1}{n^{2}(2n+1)}\,,
\nonumber\\
J_{111}^{(1)}&= & \frac{x}{2}\sum_{n=1}^{\infty}\frac{\left(\begin{array}{c}
	2n\\
	n
	\end{array}\right)}{\left(\begin{array}{c}
	4n\\
	2n
  \end{array}\right)}\,\frac{\left(-x^{2}\right)^{n}}{n(n+1)(4n+1)}\left( \ln x-S_{1}+3\oS-2\ooS-\frac{1}{2n}-\frac{1}{2(n+1)}-\frac{2}{1+4n}\right)\,,
\nonumber\\
J_{111}^{(2)}&= & \frac{1}{2}\sum_{n=1}^{\infty}\frac{\left(-16x^{2}\right)^{n}}{\left(\begin{array}{c}
	2n\\
	n
	\end{array}\right)\left(\begin{array}{c}
	4n\\
	2n
	\end{array}\right)}\,\frac{1}{n^{2}(2n-1)(2n+1)}\,.
\end{eqnarray}
Here, for brevity, we have omitted the arguments of the harmonic sums $S_1(n)$
and introduced the short-hand notations
\begin{equation}
S_1 = S_1(n-1)\,, \qquad \overline{S_1} = S_1(2n-1)\,,
\qquad \overline{\overline{S_1}} = S_1(4n-1) \,.
\end{equation}
For the practical applications mentioned in Section~\ref{sec:intro},
we also need the $\OO(\ep)$ terms of the sunset diagrams.
We do not present them here for a general value of $x$, since the
corresponding expressions are too lengthy.
However, they may be found in Ref.~\cite{sunsetsNRQCD}.

Similarly, for the three- and four-point two-loop integrals, we have\footnote{Here, we have summed series corresponding to multiple polylogarithms.}
\begin{eqnarray}
  e^{2\ep\gamma_{E}}\left(M^{2}\right)^{2\ep+1}J_{01112}^{mM}&= & -\frac{1}{2}+\frac{1}{x}\left[ \Li_{3}(x)-\frac{1}{2}\ln x\Li_{2}(x)\right] +J_{1112}^{(1)}+J_{1112}^{(2)} + \OO(\ep)\,,
  \nonumber\\
  e^{2\ep\gamma_{E}}\left(M^{2}\right)^{2\ep}J_{01111}^{mM}&= & -\frac{1}{2\ep^{2}}+\frac{1}{\ep}\left(\ln x-\frac{1}{2}\right)-\frac{3}{2}-\frac{\zeta_{2}}{2} \nonumber \\
&&{} +[2-\ln(1-x)]\ln x-\frac{\ln^{2}x}{2}-\Li_{2}(x) \nonumber \\
  &&{} +\frac{1}{x}\left\{ 2\Li_{3}(x)+\Li_{2}(x)+\ln x\left[\ln(1-x)-\Li_{2}(x)\right]\right\} \nonumber \\
&&{} +J_{1111}^{(1)}+J_{1111}^{(2)} + \OO(\ep)\,,
\nonumber\\
e^{2\ep\gamma_{E}}\left(M^{2}\right)^{2\ep+2}J_{11112}^{mM}&= & \frac{1}{2x^{3/2}}\left[ \Li_{2}\left(-\sqrt{x}\,\right)-\Li_{2}\left(\sqrt{x}\,\right)+\frac{1}{2}\ln x
  \ln\frac{1+\sqrt{x}}{1-\sqrt{x}}\right] \nonumber \\
&&{} +\frac{1}{4x}\left\{ 3+\Li_{2}(x)+\ln x\left[\ln(1-x)-2\right]\right\} +J_{11112}^{(1)}+J_{11112}^{(2)} + \OO(\ep)\,,
\nonumber\\
e^{2\ep\gamma_{E}}\left(M^{2}\right)^{2\ep+1}J_{11111}^{mM}&= & \frac{1}{x^{3/2}}\left[ 2\Li_{2}\left(-\sqrt{x}\,\right)-2\Li_{2}\left(\sqrt{x}\,\right)+\ln x
  \ln\frac{1+\sqrt{x}}{1-\sqrt{x}}\right] \nonumber \\
&&{} +\frac{1}{x}\left\{ \frac{1}{2\ep}+4-\Li_{3}(x)+\Li_{2}(x)+\ln x\left[\frac{1}{2}\Li_{2}(x)+\ln(1-x)-\frac{5}{2}\right]\right\} \nonumber  \\ &&{}+J_{11111}^{(1)}+J_{11111}^{(2)} + \OO(\ep)\,,
\end{eqnarray}
where
\begin{eqnarray}
J_{1112}^{(1)}&= & \frac{1}{8x}\sum_{n=1}^{\infty}\left(-x^{2}\right)^{n}\frac{\left(\begin{array}{c}
	2n\\
	n
	\end{array}\right)}{\left(\begin{array}{c}
	4n\\
	2n
  \end{array}\right)}\,\frac{1}{n^{2}}\left( \ln x-\frac{1}{n}-S_{1}+3\oS-2\ooS\right)\, ,
\nonumber\\
J_{1112}^{(2)}&= & -\frac{1}{2}\sum_{n=1}^{\infty}\frac{\left(-16x^{2}\right)^{n}}{\left(\begin{array}{c}
	2n\\
	n
	\end{array}\right)\left(\begin{array}{c}
	4n\\
	2n
	\end{array}\right)}\,\frac{1}{(2n+1)^{2}(4n+1)}\, ,
\nonumber\\
J_{1111}^{(1)}&= & \frac{1}{4x}\sum_{n=1}^{\infty}\frac{\left(\begin{array}{c}
	2n\\
	n
	\end{array}\right)}{\left(\begin{array}{c}
	4n\\
	2n
  \end{array}\right)}\,\frac{\left(-x^{2}\right)^{n}}{n^{2}(2n-1)}\left( \frac{1}{n}+\frac{1}{2n-1}+S_{1}-3\oS+2\ooS-\ln x\right)\, ,
\nonumber\\
J_{1111}^{(2)}&= & \frac{1}{2}\sum_{n=1}^{\infty}\frac{\left(-16x^{2}\right)^{n}}{\left(\begin{array}{c}
	2n\\
	n
	\end{array}\right)\left(\begin{array}{c}
	4n\\
	2n
	\end{array}\right)}\,\frac{1}{n(2n+1)^{2}(4n+1)}\,, 
\nonumber\\
J_{11112}^{(1)}&= & \frac{1}{8x}\sum_{n=1}^{\infty}\left(-x^{2}\right)^{n}\frac{\left(\begin{array}{c}
	2n\\
	n
	\end{array}\right)}{\left(\begin{array}{c}
	4n\\
	2n
  \end{array}\right)}\,\frac{1}{n(4n+1)}\left( \ln x-\frac{1}{2n}-\frac{2}{4n+1}-S_{1}+3\oS-2\ooS\right)\, ,
\nonumber\\
J_{11112}^{(2)}&= & \frac{1}{16x^{2}}\sum_{n=1}^{\infty}\frac{\left(-16x^{2}\right)^{n}}{\left(\begin{array}{c}
	2n\\
	n
	\end{array}\right)\left(\begin{array}{c}
	4n\\
	2n
  \end{array}\right)}\,\frac{1}{n^{2}(2n-1)}\, ,
\nonumber\\
J_{11111}^{(1)}&= & \frac{1}{8x}\sum_{n=1}^{\infty}\left(-x^{2}\right)^{n}\frac{\left(\begin{array}{c}
	2n\\
	n
	\end{array}\right)}{\left(\begin{array}{c}
	4n\\
	2n
  \end{array}\right)}\,\frac{1}{n^{2}(4n+1)}\left( \frac{1}{n}+\frac{2}{4n+1}+S_{1}-3\oS+2\ooS-\ln x\right)\, ,
\nonumber\\
J_{11111}^{(2)}&= & \frac{1}{8x^{2}}\sum_{n=1}^{\infty}\frac{\left(-16x^{2}\right)^{n}}{\left(\begin{array}{c}
	2n\\
	n
	\end{array}\right)\left(\begin{array}{c}
	4n\\
	2n
	\end{array}\right)}\,\frac{1}{n^{2}(2n-1)^{2}}\, .
\end{eqnarray}

\subsection{Integral representations}
\label{integralRep}

Although the series expansions obtained in Subsection~\ref{seriesRep} are rapidly converging, it is useful to also find integral representations for the considered diagrams. To derive integral representations, it is helpful to rewrite the above series representations in the following form:
\begin{eqnarray}
J_{122}^{(2)}&=& -\frac{1}{8x^2}  \sum_{n=1}^{\infty} \left(-16x^2\right)^n \frac{\Gamma^2(n)\Gamma(2n-1)}{\Gamma(4n-1)}\,,
\nonumber\\
J_{112}^{(2)}&=& -\frac{1}{4}  \sum_{n=1}^{\infty} \left(-16x^2\right)^n \frac{\Gamma^2(n)\Gamma(2n+1)}{\Gamma(4n+1)}
\, \frac{1}{2n+1}\,,
\nonumber\\
J_{111}^{(2)}&=& 
\sum_{n=1}^{\infty}  \left(-16x^2\right)^n \frac{\Gamma(n)\Gamma(n+1)\Gamma(2n-1)}{\Gamma(4n+1)} \, \frac{1}{2n+1}\,,
\nonumber\\
J_{1112}^{(2)}&=& \frac{1}{16x^2}
\sum_{n=1}^{\infty}  \left(-16x^2\right)^n \frac{\Gamma^2(n)\Gamma(2n-1)}{\Gamma(4n-1)} \, \frac{1}{2n-1}\,,
\nonumber\\
J_{1111}^{(2)}&=& -
\sum_{n=1}^{\infty}  \left(-16x^2\right)^n \frac{\Gamma(n)\Gamma(n+1)\Gamma(2n+1)}{\Gamma(4n+3)}\, \frac{1}{2n+1}\,,
\nonumber\\
J_{11112}^{(2)}&=& \frac{1}{8x^2}  \sum_{n=1}^{\infty} \left(-16x^2\right)^n \frac{\Gamma(n)\Gamma(n+1)\Gamma(2n-1)}{\Gamma(4n+1)}\,,
\nonumber\\
J_{11111}^{(2)}&=& -\frac{1}{4x^2} \sum_{n=1}^{\infty} \left(-16x^2\right)^n \frac{\Gamma(n))\Gamma(n+1)\Gamma(2n-1)}{\Gamma(4n+1)}
\, \frac{1}{2n-1}\,.
\end{eqnarray}
Now, we write
\begin{equation}
J_{1....}^{(1)}= \left.\frac{d}{d\delta}J_{1....}^{(1)}(\delta) \right|_{\delta=0}\,,
\end{equation}
where
\begin{eqnarray}
J_{122}^{(1)}(\delta)&=&
\frac{1}{x}  \sum_{n=1}^{\infty} \left(-x^2\right)^n  x^{-\delta}
\frac{\Gamma^2(2n+1-\delta)\Gamma(2n-\delta)}{\Gamma^2(n+1-\delta/2)\Gamma(4n+1-2\delta)} \, ,
\nonumber\\
J_{112}^{(1)}(\delta)&=& 
\frac{x}{2}  \sum_{n=0}^{\infty} \left(-x^2\right)^n x^{-\delta}
\frac{\Gamma^3(2n+1-\delta)}{\Gamma(n+1-\delta/2)\Gamma(n+2-\delta/2)\Gamma(4n+2-2\delta)}\,,
\nonumber\\
J_{111}^{(1)}(\delta)&=&   - x   \sum_{n=1}^{\infty} \left(-x^2\right)^n x^{-\delta}
\frac{\Gamma^2(2n+1-\delta)\Gamma(2n-\delta)}{\Gamma(n+1-\delta/2)\Gamma(n+2-\delta/2)\Gamma(4n+2-2\delta)}\,,
\nonumber\\
J_{1112}^{(1)}(\delta)&=& 
-\frac{1}{2x} \sum_{n=1}^{\infty} \left(-x^2\right)^n x^{-\delta}
\frac{\Gamma(2n+1-\delta)\Gamma^2(2n-\delta)}{\Gamma^2(n+1-\delta/2)\Gamma(4n+1-2\delta)}\,,
\nonumber\\
J_{1111}^{(1)}(\delta)&=&  \frac{1}{x} \sum_{n=1}^{\infty} \left(-x^2\right)^n x^{-\delta}
\frac{\Gamma(2n+1-\delta)\Gamma(2n-\delta)\Gamma(2n-1-\delta)}{\Gamma^2(n+1-\delta/2)\Gamma(4n+1-2\delta)}\,,
\nonumber\\
J_{11112}^{(1)}(\delta)&=& 
-\frac{1}{4x} \sum_{n=1}^{\infty} \left(-x^2\right)^n x^{-\delta}
\frac{\Gamma^2(2n+1-\delta)\Gamma(2n-\delta)}{\Gamma^2(n+1-\delta/2)\Gamma(4n+1-2\delta)}\,,
\nonumber\\
J_{11111}^{(1)}(\delta)&=&  \frac{1}{2x} \sum_{n=1}^{\infty} \left(-x^2\right)^n x^{-\delta}
\frac{\Gamma(2n+1-\delta)\Gamma^2(2n-\delta)}{\Gamma^2(n+1-\delta/2)\Gamma(4n+2-2\delta)}\,.
\end{eqnarray}
Starting from this point, the corresponding integral representations can be obtained in two steps. First, we rewrite the functions
$\Gamma(4n+2a(n,\delta))$, with some $a(n,\delta)$ depending on the series under consideration, in the denominators as products of two $\Gamma$ functions, $\Gamma(2n+a(n,\delta))$ and  $\Gamma(2n+a(n,\delta)+1/2)$. Second, we represent
the ratio of the function $\Gamma(2n+b(n,\delta))$, with some $b(n,\delta)$ depending on the series under consideration, in the numerator and the function
$\Gamma(2n+a(n,\delta)+1/2)$ in the denominator as
\begin{equation}
\frac{\Gamma(2n+b(n,\delta))}{\Gamma(2n+a(n,\delta)+1/2)} = \int_0^1  dt  \frac{t^{2n+b(n,\delta)-1}
	(1-t)^{a(n,\delta)-b(n,\delta)-1/2}}{\Gamma\bigl(a(n,\delta)-b(n,\delta)+1/2\bigr)}\,.
\end{equation}
The remaining series can now be summed, and we are left with the sought integral representations over $t$. This way, we obtain the following integral representations for the two-point diagrams: 
\begin{eqnarray}
  J_{122}^{(1)}&= & -\frac{1}{2x}\int_{0}^{1}\frac{dt}{t\sqrt{1-t}}\left[\frac{1}{\sqrt{1+4A^{2}}}L_{1}(A)-2\ln A\right]\,,
\\
J_{122}^{(2)}&= & -\frac{1}{2x}\int_{0}^{1}\frac{dt}{t\sqrt{1-t}}\,\frac{1}{\sqrt{1+4A^{2}}}L_{2}(A)\,,
\\
J_{112}^{(1)}&= & -\frac{1}{x}\int_{0}^{1}\frac{dt}{t^{2}\sqrt{1-t}}\left[\sqrt{1+4A^{2}}L_{1}(A)-2\ln A\right]\,,
\label{IntJ112.1}
\\
J_{112}^{(2)}&= & -\frac{1}{x}\int_{0}^{1}\frac{dt}{t^{2}\sqrt{1-t}}\left[\sqrt{1+4A^{2}}L_{2}(A)+4A\right]\,,
\label{IntJ112.2}\\
J_{111}^{(1)}&= & \frac{4}{x}\int_{0}^{1}dt\frac{\sqrt{1-t}}{t^{3}}\left[\sqrt{1+4A^{2}}L_{1}(A)-2\ln A+2A^{2}-4A^{2}\ln A\right]\,,
\\
J_{111}^{(2)}&= & \frac{4}{x}\int_{0}^{1}dt\frac{\sqrt{1-t}}{t^{3}}\left[\sqrt{1+4A^{2}}L_{2}(A)+4A\right]\,,
\label{IntJ111.2}
\end{eqnarray}
where $A={xt}/{4}$,
\begin{equation}
L_{1}(A)= \ln\frac{\sqrt{1+4A^{2}}-1}{\sqrt{1+4A^{2}}+1}\,,\qquad
L_{2}(A)= \ln\frac{\sqrt{1+4A^{2}}-2A}{\sqrt{1+4A^{2}}+2A}\,.
\end{equation}
Notice that the above integral representations for the sunset diagrams are simpler than those in Ref.~\cite{sunsetsNRQCD}, which involved some dilogarithms. Similarly, for the functions entering the expressions for the three- and four-point integrals, we have 
\begin{eqnarray}
  J_{1112}^{(1)}&= & -\frac{1}{4x}\int_{0}^{1}\frac{dt}{t\sqrt{1-t}}\left[\ln^{2}A-\frac{1}{4}L_{1}(A)^{2}\right]\,,
\nonumber\\
J_{1112}^{(2)}&= & \frac{1}{2x}\int_{0}^{1}\frac{dt}{t\sqrt{1-t}}\left[I_{1}(A)+2A\right]\,,
\nonumber\\
J_{1111}^{(1)}&= & \frac{1}{x}\int_{0}^{1}dt\frac{\sqrt{1-t}}{t^{2}}\left[\ln^{2}A-\frac{1}{4}L_{1}(A)^{2}\right]\,,
\nonumber\\
J_{1111}^{(2)}&= & -\frac{2}{x}\int_{0}^{1}dt\frac{\sqrt{1-t}}{t^{2}}\left[I_{1}(A)-1+2A\right]\,,
\nonumber\\
J_{11112}^{(1)}&= & \frac{1}{8x}\int_{0}^{1}dt\frac{\sqrt{1-t}}{t}\left[\frac{1}{\sqrt{1+4A^{2}}}L_{1}(A)-2\ln A\right]\,,
\nonumber\\
J_{11112}^{(2)}&= & \frac{1}{8x}\int_{0}^{1}dt\frac{\sqrt{1-t}}{t}\,\frac{1}{\sqrt{1+4A^{2}}}L_{2}(A)\,,
\nonumber\\
J_{11111}^{(1)}&= & \frac{1}{4x}\int_{0}^{1}dt\frac{\sqrt{1-t}}{t}\left[\ln^{2}A-\frac{1}{4}L_{1}(A)^{2}\right]\,,
\nonumber\\
J_{11111}^{(2)}&= & -\frac{1}{2x}\int_{0}^{1}dt\frac{\sqrt{1-t}}{t}I_{1}(A)\,,
\end{eqnarray} 
where $I_{1}(A)$ is defined in Eq.~\eqref{I1x}. Such correspondence reveals a deep relation between the subintegral expressions of the two-loop integrals and the corresponding one-loop results, in agreement with the general prescription in Eq.~\eqref{2loopA1}.


To check that the above integral representations are simplest, it is convenient to use the following relations between the functions
$J_{...12}^{(i)}$ and $J_{...11}^{(i)}$ $(i=1,2)$:
\begin{eqnarray}
\frac{d}{dx} \left(\frac{1}{x} \, J_{111}^{(i)}\right) &=& -\frac{2}{x^2} \left[ J_{112}^{(i)} -  \delta^i_1   \frac{1}{4}x(5-2\ln x) \right] 
\, ,
\label{rela.1}\\
x\frac{d}{dx} J_{1111}^{(i)} &=& -2 J_{112}^{(i)} + \delta^i_2 \, ,
\\
\frac{d}{dx} \left(x \, J_{11111}^{(i)}\right) &=& -2  J_{11112}^{(i)} \, ,
\end{eqnarray}
where $\delta^i_2$ is the Kronecker symbol, which directly follow from the corresponding series representations. 

Indeed, in the simplest forms of the integral representations, these relations can easily be reproduced. As a first example, let us consider Eq.~\eqref{rela.1} with $i=2$.
From  Eq.~\eqref{IntJ111.2}, we have
\begin{eqnarray}
\frac{d}{dx} \left(\frac{1}{x} J_{111}^{(2)}\right) &=& \frac{d}{dx} 
\int_{0}^{1}dt\frac{\sqrt{1-t}}{t} \, \frac{1}{4A^{2}} \left[\sqrt{1+4A^{2}}L_{2}(A)+4A\right] \nonumber\\
&=& \int_{0}^{1}dt\frac{\sqrt{1-t}}{t} \, \frac{d}{dx} \left\{\frac{1}{4A^{2}}  \left[\sqrt{1+4A^{2}}L_{2}(A)+4A\right]\right\}\,.
\label{rela.1.1}
\end{eqnarray}
Now, notice that the expression in brackets on the r.h.s.\ of Eq.~\eqref{rela.1.1} only depends on $A$. So, we can replace the derivative with respect to $x$ acting on it with
\begin{equation}
\frac{d}{dx} =  \frac{dA}{dx} \, \frac{d}{dA} = \frac{t}{2} \, \frac{d}{dA}\,.
\end{equation}
Similarly, we have
\begin{equation}
\frac{d}{dt} =  \frac{dA}{dt} \, \frac{d}{dA} = \frac{x}{2} \, \frac{d}{dA}\,,
\end{equation}
so that, on the r.h.s.\ of Eq.~\eqref{rela.1.1}, we can make the replacement
\begin{equation}
\frac{d}{dx} = \frac{t}{x} \, \frac{d}{dt} \, ,
\end{equation}
that is
\begin{equation}
\frac{d}{dx} \left(\frac{1}{x} J_{111}^{(2)}\right)= \int_{0}^{1}dt\frac{\sqrt{1-t}}{x} \, \frac{d}{dt}
\left\{\frac{1}{4A^{2}}  \left[\sqrt{1+4A^{2}}L_{2}(A)+4A\right]\right\} \, .
\end{equation}
Next, using IBP, we have
\begin{eqnarray}
\frac{d}{dx} \left(\frac{1}{x} J_{111}^{(2)}\right) &=&  \frac{1}{x} \left\{\sqrt{1-t} \left[\frac{1}{4A^{2}} 
\left(\sqrt{1+4A^{2}}L_{2}(A)+4A\right)\right]^{t=1}_{t=0}\right. \nonumber \\
&&{} -\left.  \int_{0}^{1} dt \left(\frac{d}{dt} \sqrt{1-t}\right)   \left[\frac{1}{4A^{2}}\left(\sqrt{1+4A^{2}}L_{2}(A)+4A\right)\right] \right\}\, .
\label{rela.1.5}
\end{eqnarray}
The first term on the r.h.s.\ of Eq.~\eqref{rela.1.5} is equal to zero, and
\begin{equation}
\frac{d}{dt} \sqrt{1-t}= - \frac{1}{2\sqrt{1-t}} \, .
\end{equation}
So, finally, we have
\begin{eqnarray}
\frac{d}{dx} \left(\frac{1}{x} J_{111}^{(2)}\right) &=& \frac{1}{2x} \int_{0}^{1} \frac{dt}{\sqrt{1-t}} 
\left\{\frac{1}{4A^{2}} \left[\sqrt{1+4A^{2}}L_{2}(A)+4A\right]\right\} \nonumber \\
&=&\frac{2}{x^3}  \int_{0}^{1} \frac{dt}{t^2\sqrt{1-t}}  \left[\sqrt{1+4A^{2}}L_{2}(A)+4A\right]
\nonumber\\
&=&  -2 \frac{1}{x^2} J_{112}^{(2)} \, ,
\label{rela.1.6}
\end{eqnarray}
where we have used Eq.~\eqref{IntJ112.2} in the third equality.
We thus observe that Eq.~\eqref{rela.1} with $i=2$ is correct.

As a second example, let us consider Eq.~\eqref{rela.1} with $i=1$.
Repeating steps similar to those in Eqs.~\eqref{rela.1.1}--\eqref{rela.1.6}, we have
  \begin{equation}
  \frac{d}{dx} \left(\frac{1}{x} J_{111}^{(1)}\right)
=  \frac{2}{x^3} \int_{0}^{1} \frac{dt}{t^2\sqrt{1-t}} \left[\sqrt{1+4A^{2}}L_{1}(A)-2\ln A+2A^{2}-4A^{2}\ln A\right] \, .
\label{rela.1.7}
\end{equation}
  Then, the evaluation of the terms proportional to $A^2$ on the r.h.s.\ of
  Eq.~\eqref{rela.1.7} gives
\begin{equation}
\frac{1}{4x} \int_{0}^{1} \frac{dt}{\sqrt{1-t}} (1-2\ln A) = \frac{1}{2x} (5-2\ln x)\,,
\end{equation}
so that, finally, we have
  \begin{eqnarray}
  \frac{d}{dx} \left(\frac{1}{x} J_{111}^{(1)}\right)
  &=&
  \frac{2}{x^3}\left\{ \int_{0}^{1} \frac{dt}{t^2\sqrt{1-t}} \left[\sqrt{1+4A^{2}}L_{1}(A)-2\ln A\right] + \frac{x^2}{4}(5-2\ln (x)
    \right\}\, \nonumber \\
&=&  -2 \frac{1}{x^2} \left[ J_{112}^{(1)} - \frac{x}{4}(5-2\ln x)    \right] \,,
  \end{eqnarray}
where we have used Eq.~\eqref{IntJ112.1} in the second equality. 
We thus observe that Eq.~\eqref{rela.1} with $i=1$ is also correct.

\subsection{Representations in terms of generalized hypergeometric functions}
\label{hypergeomRep}

For the two-point sunset-type diagrams, the results in terms of hypergeometric functions were already given in Eq.~\eqref{eq:gehyfu}. So, here we
present such results for the three- and four-point diagrams only. The series representations obtained in Subsection~\ref{seriesRep} can be readily expressed in terms of hypergeometric functions and their derivatives. We thus obtain
\begin{eqnarray}
  \lefteqn{e^{2\ep\gamma_{E}}\left(M^{2}\right)^{2\ep+1}J_{1112}= \frac{1}{x}\left[ \Li_{3}(x)-\frac{1}{2}\ln x\Li_{2}(x)\right]}
  \nonumber\\
  &&{}+\frac{1}{4}\,\frac{d}{d\delta}\left[ \frac{1}{(2\delta-1)}{}_{4}F_{3}\left(\begin{array}{c}
1,1,1,\frac{1}{2}-\delta\nonumber \\
\frac{3}{4},\frac{5}{4},\frac{3}{2}-\delta
\end{array}\Bigg|-\frac{x^{2}}{4}\right)\right]_{\delta=0}\\
&&{}+\frac{1}{4}\,\frac{d^{2}}{d\delta_{1}d\delta_{2}}\left[\frac{x^{1+\delta_{1}}}{(3+2\delta_{1})(2+\delta_{1}-2\delta_{2})}{}_{4}F_{3}\left(\begin{array}{c}
1,\frac{3}{2},\frac{3}{2}+\delta_{1},1+\frac{\delta_{1}}{2}-\delta_{2}\\
\frac{5}{4}+\frac{\delta_{1}}{2},\frac{7}{4}+\frac{\delta_{1}}{2},2+\frac{\delta_{1}}{2}-\delta_{2}
  \end{array}\Bigg|-\frac{x^{2}}{4}\right)\right]_{\delta_{1}=\delta_{2}=0} \nonumber\\
  &&{}+\OO(\ep)\,,
\nonumber\\
\lefteqn{e^{2\ep\gamma_{E}}\left(M^{2}\right)^{2\ep}J_{1111}=  -\frac{1}{2\ep^{2}}+\frac{1}{2\ep}(-1+2\ln x)+\frac{1}{2}[-3-\zeta_{2}+4\ln x-2\ln x\ln(1-x)}
\nonumber\\ &&{}-\ln^{2}x-2\Li_{2}(x)] 
 +\frac{1}{x}\left[\ln x\ln(1-x)+\Li_{2}(x)-\ln x\Li_{2}(x)+2\Li_{3}(x)\right]\nonumber \\
&&{} +\frac{x^{2}}{15}\,\frac{d}{d\delta}\left[ \frac{1}{(2\delta-3)}{}_{4}F_{3}\left(\begin{array}{c}
1,1,2,\frac{3}{2}-\delta\\
\frac{7}{4},\frac{9}{4},\frac{5}{2}-\delta
\end{array}\Bigg|-\frac{x^{2}}{4}\right)\right]_{\delta=0}\nonumber\\
&&{} +\frac{1}{2}\,\frac{d^{2}}{d\delta_{1}d\delta_{2}}\left[ \frac{x^{1-\delta_{1}}(\delta_{1}-1)}{(8\delta_{1}-3)(\delta_{1}+2\delta_{2}-2)}{}_{4}F_{3}\left(\begin{array}{c}
1,\frac{1}{2}-\frac{\delta_{1}}{2},\frac{3}{2}-\frac{\delta_{1}}{2},1-\frac{\delta_{1}}{2}-\delta_{2}\\
\frac{5}{4}-\frac{\delta_{1}}{2},\frac{7}{4}-\frac{\delta_{1}}{2},2-\frac{\delta_{1}}{2}-\delta_{2}
 \end{array}\Bigg|-\frac{x^{2}}{4}\right)\right]_{\delta_{1}=\delta_{2}=0} \nonumber\\
  &&{}+\OO(\ep)\,,
\nonumber\\
\lefteqn{e^{2\ep\gamma_{E}}\left(M^{2}\right)^{2\ep+2}J_{11112}=  \frac{1}{4x}\left[3-2\ln x+\ln x\ln(1-x)+\Li_{2}(x)\right]}\nonumber\\
&&{} +\frac{1}{4x^{3/2}}\left[\ln x\ln\frac{1+\sqrt{x}}{1-\sqrt{x}}+2\Li_{2}(-\sqrt{x}\,)-2\Li_{2}(\sqrt{x}\,)\right]
-\frac{1}{12}{}_{4}F_{3}\left(\begin{array}{c}
\frac{1}{2},1,1,1\\
\frac{5}{4},\frac{3}{2},\frac{7}{4}
\end{array}\Bigg|-\frac{x^{2}}{4}\right)\nonumber\\
&&{} -\frac{1}{4}\,\frac{d}{d\delta}\left[\frac{x^{1+\delta}(1+\delta)^{2}}{(2+\delta)(15+46\delta)}{}_{4}F_{3}\left(\begin{array}{c}
1,1+\frac{\delta}{2},\frac{3}{2}+\frac{\delta}{2},\frac{3}{2}+\frac{\delta}{2}\\
\frac{7}{4}+\frac{\delta}{2},2+\frac{\delta}{2},\frac{9}{4}+\frac{\delta}{2}
\end{array}\Bigg|-\frac{x^{2}}{4}\right)\right]_{\delta=0}+\OO(\ep)\,,
\nonumber\\
\lefteqn{e^{2\ep\gamma_{E}}\left(M^{2}\right)^{2\ep+1}J_{11111}= \frac{1}{x}\left\{ \frac{1}{2\ep}+\frac{1}{2}\left[8-5\ln x+2\ln x\ln(1-x)+2\Li_{2}(x)+\ln x\Li_{2}(x)\right.\right.}
\nonumber \\
&&{}-\left.
\vphantom{\frac{1}{2\ep}}
\left.2\Li_{3}(x)\right]\right\}
+\frac{1}{x^{3/2}}\left[\ln x\ln\frac{1+\sqrt{x}}{1-\sqrt{x}}+2\Li_{2}\left(-\sqrt{x}\,\right)-2\Li_{2}\left(\sqrt{x}\,\right)\right]\nonumber\\
&&{} -\frac{1}{12}\,\frac{d}{d\delta}\left[\frac{1}{(2\delta-1)}{}_{4}F_{3}\left(\begin{array}{c}
1,1,1,\frac{1}{2}-\delta\\
\frac{5}{4},\frac{7}{4},\frac{3}{2}-\delta
\end{array}|-\frac{x^{2}}{4}\right)\right]_{\delta=0}\nonumber\\
&&{} -\frac{1}{4}\,\frac{d^{2}}{d\delta_{1}d\delta_{2}}\left[\frac{x^{1-\delta_{1}}(1-\delta_{1})^{2}}{(15-46\delta_{1})(2-\delta_{1}-2\delta_{2})}{}_{4}F_{3}\left(\begin{array}{c}
1,\frac{3}{2}-\frac{\delta_{1}}{2},\frac{3}{2}-\frac{\delta_{1}}{2},1-\frac{\delta_{1}}{2}-\delta_{2}\\
\frac{7}{4}-\frac{\delta_{1}}{2},\frac{9}{4}-\frac{\delta_{1}}{2},2-\frac{\delta_{1}}{2}-\delta_{2}
  \end{array}\Bigg|-\frac{x^{2}}{4}\right)\right]_{\delta_{1}=\delta_{2}=0} \nonumber\\
  &&{} +\OO(\ep)\,.
\end{eqnarray}

\subsection{Equal-mass case}
\label{equalmass}

In Ref.~\cite{sunsetsNRQCD}, we showed that, in the equal-mass case, the results for the two-loop sunset diagrams with the special kinematics in Eq.~\eqref{kinematics} can be written in terms of five ``elliptic" sums,
\begin{equation}
 \sum_{n=1}^{\infty}
 \frac{(-16)^n}{\left(2n \atop n\right)\left(4n \atop 2n\right)}
 \left\{ 1; \frac{1}{n} \right\}\,,\qquad
  \sum_{n=1}^{\infty} (-1)^n
  \frac{\left(2n \atop n\right)}{\left(4n \atop 2n\right)}\left\{
\phi ; \frac{\phi}{n}; \frac{1}{n^2}  	
\right\}\,,
\end{equation} 
where $\phi = S_1 - 3\oS + 2\ooS$.
Specifically, up to $\OO (\ep)$ terms, we have\footnote{In the equal-mass case, only two of the sunset diagrams correspond to master integrals, namely
$J_{111}^{mm}$ and $J_{112}^{mm}$.} 
\begin{eqnarray}
J_{111}^{mm} &=& -\frac{3}{2\ep^2} - \frac{19}{4\ep}
+ \Sigma_{111} +
+ \OO(\ep) \,, \nonumber \\
J_{112}^{mm} &=& -\frac{1}{2\ep^2} - \frac{1}{2\ep}
+ \Sigma_{112} 
+ \OO(\ep) \,,
\end{eqnarray}
where 
\begin{eqnarray}
\Sigma_{111} & = &
- \frac{215}{24}  + \frac{9}{4}\zeta_2 \nonumber\\
&&{} + \sum_{n=1}^{\infty} (-1)^n
\frac{\left(2n \atop n\right)}{\left(4n \atop 2n\right)}
\left( - \frac{5}{2}\phi + \frac{15\phi}{4n} + \frac{15}{8n^2}
\right)
+ \sum_{n=1}^{\infty}
\frac{(-16)^n}{\left(2n \atop n\right)\left(4n \atop 2n\right)}
\left( \frac{25}{3} - \frac{25}{6n} \right)
\nonumber\\
& = & -9.03056576107922587907436223954936770213033473413\dots \,,
\nonumber\\
\Sigma_{112} & = &
\frac{29}{18} + \frac{3}{4}\zeta_2 \nonumber\\
&&{} + \sum_{n=1}^{\infty} (-1)^n
\frac{\left(2n \atop n\right)}{\left(4n \atop 2n\right)}
\left( - \frac{5}{2}\phi + \frac{5\phi}{4n} + \frac{5}{8n^2}
\right)
+ \sum_{n=1}^{\infty}
\frac{(-16)^n}{\left(2n \atop n\right)\left(4n \atop 2n\right)}
\left( \frac{50}{9} - \frac{35}{18n} \right)
\nonumber\\
& = & 0.0113804720812563731826135489564394881100859890024139\dots \,.
\end{eqnarray}
Two of the ``elliptic" sums above were shown to be expressible in terms of elliptic integrals of the first and second kinds. Namely, we have
\begin{eqnarray}
\sum_{n=1}^{\infty} (-1)^n
\frac{\left(2n \atop n\right)}{\left(4n \atop 2n\right)}
&=&  -\frac{1}{2} + \frac{1}{10}p
+ \frac{1}{5}p e_1 - \frac{1}{10}p f_1 \,,
\nonumber\\
\sum_{n=1}^{\infty} \frac{(-1)^n}{n}\,
\frac{\left(2n \atop n\right)}{\left(4n \atop 2n\right)}
&=& - \frac{4}{3}\ln2 +\frac{1}{3}f_1 \,,
\end{eqnarray}
where 
\begin{eqnarray}
f_1 &=& \frac{1}{\sqrt{p}}
F\left( 2 \arctan\sqrt{p}, \frac{1+p}{2p}\right)
= \frac{1}{\sqrt{p}}
F\left( 2 \varphi , \frac{1}{2\sin^2\varphi}\right)
= 1.8829167613\dots \,,
\nonumber\\
e_1 &=& \frac{1}{\sqrt{p}}
E\left( 2 \arctan\sqrt{p},\, \frac{1+p}{2p}\right)
= \frac{1}{\sqrt{p}}
E\left( 2 \varphi ,\, \frac{1}{2\sin^2\varphi}\right)
= 0.9671227369\dots  \,,
\end{eqnarray}
with $p = \sqrt{5}$ and $\varphi = \arctan\sqrt{p}$.
Here, $F$ and $E$ are the elliptic integrals of the first and second kinds, respectively. They are defined as
\begin{equation}
F(\phi,k)=\int_0^\phi{\rm d}\theta (1-k\sin^2\theta)^{-1/2}\,,
\qquad
E(\phi,k)=\int_0^\phi{\rm d}\theta (1-k\sin^2\theta)^{1/2}\,.
\end{equation}

However, to write equal-mass expressions for the three- and four-point functions, we need three additional ``elliptic" sums,
\begin{equation}
\sum_{n=1}^{\infty}
\frac{(-16)^n}{\left(2n \atop n\right)\left(4n \atop 2n\right)}\,
\frac{1}{(2n+1)^2}\,,\qquad
\sum_{n=1}^{\infty} (-1)^n
\frac{\left(2n \atop n\right)}{\left(4n \atop 2n\right)}\left\{
 \frac{\phi}{n^2}; \frac{1}{n^3}  	
\right\}\,.
\end{equation} 
Then, the expressions for the three- and four-point functions are given by
\begin{eqnarray}
J_{1111}^{mm} &=& -\frac{1}{2\ep^2} - \frac{1}{2\ep}
+ \Sigma_{1111}
+ \OO(\ep) \,,
\nonumber\\
J_{1112}^{mm} &=& \Sigma_{1112} + \OO(\ep) \,,
\nonumber\\
J_{11111}^{mm} &=&  \frac{1}{2\ep}
+ \Sigma_{11111}
+ \OO(\ep) \,,
\nonumber\\
J_{11112}^{mm} &=& \Sigma_{11112} + \OO(\ep) \,,
\end{eqnarray}
where
\begin{eqnarray}
\Sigma_{1111} & = &
\frac{23}{6}  - \frac{11}{4}\zeta_2 + 2\zeta_3
 + \sum_{n=1}^{\infty} (-1)^n
\frac{\left(2n \atop n\right)}{\left(4n \atop 2n\right)}
\left( - \frac{5}{2}\phi - \frac{5\phi}{4n} - \frac{5}{8n^2} -\frac{\phi}{4n^2} -\frac{1}{n^3}
\right) \nonumber \\
&&{} + \sum_{n=1}^{\infty}
\frac{(-16)^n}{\left(2n \atop n\right)\left(4n \atop 2n\right)}
\left( \frac{5}{3} + \frac{7}{6n} + \frac{1}{(2n+1)^2}\right)
\nonumber\\
& = & -0.15218784758802924439663920716574611558826615862870\dots \,,
\nonumber\\
\Sigma_{1112} & = &
\frac{17}{18}  - \frac{1}{2}\zeta_2 + \zeta3
+ \sum_{n=1}^{\infty} (-1)^n
\frac{\left(2n \atop n\right)}{\left(4n \atop 2n\right)}
\left( -\frac{\phi}{8n^2} -\frac{1}{8n^3}
\right) \nonumber \\
&&{} + \sum_{n=1}^{\infty}
\frac{(-16)^n}{\left(2n \atop n\right)\left(4n \atop 2n\right)}
\left( -\frac{5}{9} + \frac{4}{9n} + \frac{1}{2(2n+1)^2}\right)
\nonumber\\
& = & 0.78220975700922928331809331100901619764008851521687\dots \,,
\nonumber\\
\Sigma_{11111} & = &
\frac{1}{6}  + \zeta_2 - \zeta_3
+ \sum_{n=1}^{\infty} (-1)^n
\frac{\left(2n \atop n\right)}{\left(4n \atop 2n\right)}
\left( \frac{5\phi}{2n} + \frac{5}{4n^2} +\frac{\phi}{8n^2} +\frac{1}{8n^3}
\right) \nonumber \\
&&{} + \sum_{n=1}^{\infty}
\frac{(-16)^n}{\left(2n \atop n\right)\left(4n \atop 2n\right)}
\left( \frac{10}{3} - \frac{8}{3n} - \frac{1}{2(2n+1)^2}\right)
\nonumber\\
& = & -0.34409005701793797204748426616084857709945913956561\dots \,,
\nonumber\\
\Sigma_{11112} & = &
-\frac{1}{9}  + \frac{1}{8}\zeta_2
+ \sum_{n=1}^{\infty} (-1)^n
\frac{\left(2n \atop n\right)}{\left(4n \atop 2n\right)}
\left( \frac{5\phi}{8n} + \frac{5}{16n^2}
\right)  + \sum_{n=1}^{\infty}
\frac{(-16)^n}{\left(2n \atop n\right)\left(4n \atop 2n\right)}
\left( \frac{25}{36} - \frac{5}{9n}\right)
\nonumber\\
& = & -0.14047007500217717218234773878795809486484265608719\dots \,.
\end{eqnarray}

We are dealing here with incomplete elliptic integrals; see Section~6 of
Ref.~\cite{Ablinger:2013eba} for more details.

\section{Conclusions}
\label{conclusion}

In this paper, we considered a class of two-loop diagrams with two different masses, $m$ and $M$, in the special kinematic regime defined by Eq.~\eqref{kinematics}, which play a crucial role in nonrelativistic effective field theories
like NRQCD and NRQED.
For all these diagrams, we presented analytic results in three different forms.
First, we found explicit expressions for the coefficients of the series expansions in mass ratio $x=m^2/M^2$. These allow for the numerical restoration of the results at finite values of $x$ using, for example, Pad\'e approximants. Second, we provided integral representations.
These should allow us to express the obtained results also in term of elliptic polylogarithms \cite{Beilinson:1994,Wildeshaus,Levin:1997,Levin:2007,Enriquez:2010,Brown:2011,Bloch:2013tra,Adams:2014vja,Bloch:2014qca,Adams:2015gva,Adams:2015ydq,Adams:2016xah,Remiddi:2017har,Broedel:2017kkb,Broedel:2017siw,Broedel:2018iwv,Broedel:2018qkq,Broedel:2019hyg,Broedel:2019tlz,Bogner:2019lfa,Broedel:2019kmn}, which will be the subject of one of our subsequent publications.
Third, we also found representations in terms of generalized hypergeometric
functions, up to terms of $\mathcal{O}(\ep)$.
Following Ref.~\cite{KalmykovKniehl}, we plan to find the exact results for the considered Feynman diagrams without performing $\ep$ expansions. For $x=1$, our results could be expressed in terms of fast-converging, alternating series. The latter, in general, satisfy a set of relations, which can be found, for example, with the help of the PSLQ algorithm \cite{PSLQ}. We were able to express some of these sums in terms of elliptic integrals of the first and second kinds. In general, we expect relations between our sums and elliptic multiple zeta values \cite{Enriquez:2010,eMZV2,eMZV3,eMZV4,eMZV5,eMZV7,eMZV6}, which we leave for future work.

\section*{Acknowledgments}

The authors are grateful to M.~Yu.~Kalmykov and R.~N.~Lee for fruitful discussions and valuable comments.
The work of B.A.K. and O.L.V. was supported in part by the German Federal Ministry for Education and Research (BMBF) through Grant No.\ 05H18GUCC1 and by the
German Research Foundation (DFG) through Grant No.\ KN~365/12-1.
The work of A.V.K. was supported in part by the Russian Foundation for Basic Research (RFBR) through Grant No.\ 17-02-00872, by the Russian Ministry of Science and Higher Education under Contract No.\ 02.A03.21.0003 as of 27.08.2013, and by
the Heisenberg-Landau program. The work of A.I.O. was supported in part by the RFBR through Grant No.\ 17-02-00872 and by the Foundation for the Advancement of Theoretical Physics and Mathematics ``BASIS".

\appendix

\section{Frobenius solution for a system of differential equations}
\label{diffeqFrobenius}

To use IBP and write down a system of differential equations for the integrals under consideration, it is convenient to introduce somewhat more general two-loop integrals, 
\begin{eqnarray}
I_{a_{1},\ldots,a_{7}}&=&\int\frac{d^{d}kd^{d}l}{\pi^{d}}\,\frac{1}{(k^{2})^{a_{1}}\left[(k+q_{1})^{2}\right]^{a_{2}}\left[(l+q_{1})^{2}-m^{2}\right]^{a_{3}}\left[(l-q_{2})^{2}-m^{2}\right]^{a_{4}}\left[l^{2}-m^{2}\right]^{a_{5}}}\nonumber\\
&&{}\times \frac{1}{\left[(k-l)^{2}-M^{2}\right]^{a_{6}}\left[(k+\frac{q_{1}}{2}-\frac{q_{2}}{2})^{2}-M^{2}\right]^{a_{7}}}\, ,
\end{eqnarray}
so that 
\begin{equation}
I_{00a_{3}a_{4}a_{5}a_{6}a_{7}} =J_{a_{3}a_{4}a_{5}a_{6}a_{7}}^{mM}\,.
\end{equation}
Now, introducing the vector of master integrals, 
\begin{equation}
\mathbf{I=}\left(\begin{array}{c}
I_{0000011}\\
I_{0000101}\\
I_{0000111}\\
I_{0000121}\\
I_{0000211}\\
I_{0001011}\\
I_{0001111}\\
I_{0002011}\\
I_{0010111}\\
I_{0011101}
\end{array}\right)\,,
\label{mastersvector}
\end{equation}
and using IBP, we obtain the following system of differential equations:\footnote{We used the program package {\tt LiteRed} \cite{Litered1,Litered2} for this purpose.}
\begin{equation}
  \frac{d\,\mathbf{I}}{dx}  =\mathbf{M}\,\mathbf{I}\,,
  \label{eq:dimi}
\end{equation}
where
\begin{gather}
\mathbf{M}=\left(\begin{array}{cccccccccc}
		\frac{2(\ep-1)}{x} & 0 & 0 & 0 & 0 & 0 & 0 & 0 & 0 & 0\\
		0 & \frac{\ep-1}{x} & 0 & 0 & 0 & 0 & 0 & 0 & 0 & 0\\
		0 & 0 & 0 & -\frac{2}{x^{2}} & 0 & 0 & 0 & 0 & 0 & 0\\
		\frac{(\ep-1)^{2}x(x+2)}{2(x^{2}+4)} & -\frac{2(\ep-1)^{2}x}{x^{2}+4} & \frac{(2\ep-1)(3\ep-2)(x+2)}{2(x^{2}+4)} & \frac{-6+16\ep-x+2\ep x+\ep x^{2}}{x(x^{2}+4)} & \frac{(2\ep-1)x}{x^{2}+4} & 0 & 0 & 0 & 0 & 0\\
		-\frac{(\ep-1)^{2}x}{x^{2}+4} & \frac{(\ep-1)^{2}(x+2)}{x^{2}+4} & -\frac{(2\ep-1)(3\ep-2)}{x^{2}+4} & \frac{(2\ep-1)(x-4)}{x(x^{2}+4)} & -\frac{(2\ep-1)(x-2)}{x(x^{2}+4)} & 0 & 0 & 0 & 0 & 0\\
		-\frac{(\ep-1)^{2}}{2\ep-1} & 0 & 0 & 0 & 0 & \frac{3\ep-2}{x} & 0 & \frac{2}{x} & 0 & 0\\
		-\frac{(\ep-1)^{2}}{2(2\ep-1)} & 0 & \frac{3\ep-2}{2x} & \frac{1}{x^{2}} & \frac{1}{x} & 0 & \frac{\ep-1}{x} & \frac{1}{x} & 0 & 0\\
		\frac{3(\ep-1)^{2}x}{4(x-1)} & -\frac{(\ep-1)^{2}}{2(x-1)} & 0 & 0 & 0 & -\frac{(2\ep-1)(3\ep-2)}{4(x-1)} & 0 & -\frac{(2\ep-1)(x+1)}{2x(x-1)} & 0 & 0\\
		-\frac{(\ep-1)^{2}}{2(2\ep-1)} & 0 & \frac{3\ep-2}{2x} & \frac{1}{x^{2}} & \frac{1}{x} & 0 & 0 & \frac{1}{x} & \frac{\ep-1}{x} & 0\\
		0 & 0 & 0 & 0 & 0 & 0 & 0 & 0 & 0 & \frac{\ep-1}{x}
\end{array}\right)\,.
\end{gather}

We seek for the solution of Eq.~\eqref{eq:dimi} using the Frobenius method as presented in Ref.~\cite{BarkatouFrobenius}.\footnote{For other presentations of the Frobenius method appropriate to linear systems of differential equations, see Refs.~\cite{sixroot7,Frobenius1,Frobenius2,Frobenius3,Frobenius4}.} To this end, we first need to reduce Eq.~\eqref{eq:dimi} to Fuchsian form, i.e.\
to obtain an equivalent system that only has first-order poles in the variable
$x$ as $x\to0$. The required transformation matrix is given by\footnote{It was found with the program package {\tt Fuchsia} \cite{Fuchsia} and then checked with the program package {\tt Libra} \cite{Libra}.} 
\begin{equation}
\frac{d\,\mathbf{\widetilde{I}}}{dx} = \mathbf{\widetilde{M}}\,
  \mathbf{\widetilde{I}}\,,\qquad
\mathbf{I}=\mathbf{T}\,\mathbf{\widetilde{I}}\,,
\end{equation}
where\\
\scalebox{0.97}{\parbox{.5\linewidth}{%
\begin{eqnarray}
\mathbf{T}= \left(\begin{array}{cccccccccc}
1 & 0 & 0 & 0 & 0 & 0 & 0 & 0 & 0 & 0\\
0 & 1 & 0 & 0 & 0 & 0 & 0 & 0 & 0 & 0\\
0 & 0 & -\frac{-2i-(1-2i)x+x^{2}}{x} & 0 & 0 & 0 & 0 & 0 & 0 & 0\\
0 & 0 & 0 & 1-x & 0 & 0 & 0 & 0 & 0 & 0\\
0 & 0 & 0 & 0 & 1-x & 0 & 0 & 0 & 0 & 0\\
0 & 0 & 0 & 0 & 0 & 1-x & 0 & 0 & 0 & 0\\
0 & 0 & 0 & 0 & 0 & 0 & -\frac{-2i-(1-2i)x+x^{2}}{x} & 0 & 0 & 0\\
\frac{3(1-2\ep+\ep^{2})x}{4(2\ep-1)} & -\frac{(1-2\ep+\ep^{2})x}{2(2\ep-1)} & 0 & 0 & 0 & 0 & 0 & 1-x & 0 & 0\\
0 & 0 & 0 & 0 & 0 & 0 & -\frac{-2i+2ix+x^{2}}{x} & 0 & 1 & 0\\
0 & 0 & 0 & 0 & 0 & 0 & 0 & 0 & 0 & 1
\end{array}\right)\,.
\nonumber\\
&&
\end{eqnarray}
}}

For the reader's convenience, we use the notation of Ref.~\cite{BarkatouFrobenius} in what follows. Then, the system we wish to solve is rewritten as\footnote{When passing to the representation of the system in terms of the $\mathbf{A}_i$ matrices, we multiply our original system by the common denominator of the $\mathbf{\widetilde{M}}$ matrix, so that the $\mathbf{A}_i$ matrices become polynomials in $x$.}
\begin{equation}
{\mathcal{L}}(\mathbf{\widetilde{I}}) = \mathbf{A}_1 (x) x\frac{d\mathbf{\widetilde{I}}}{dx} + \mathbf{A}_0 (x) \mathbf{\widetilde{I}} = \mathbf{0}  = x\frac{d \mathbf{\widetilde{I}}}{dx} - x \mathbf{\widetilde{M}}\,\mathbf{\widetilde{I}}\, , \label{BarkatouSystem}
\end{equation}
where $\mathbf{A}_i (x) = \sum_{j=0}^{\infty} \mathbf{A}_{i,j} x^j\in \mathbb{C}[x]^{n\times n}$
with $i=0,1$ and $n=10$, and $\mathbf{A}_{i,j}$ are equal to zero for sufficiently large
values of $j$. 

Now, following Barkatou and Gluzeau \cite{BarkatouFrobenius} and starting from the system in Eq.~\eqref{BarkatouSystem}, we consider a more general inhomogeneous system,
\begin{equation}
\mathcal{L} (\mathbf{\widetilde{I}}) = \mathbf{L}_0(\lambda)\mathbf{g}_0 (\lambda) x^{\lambda}\, , \label{nonhom-resonance-free}
\end{equation} 
where $\mathbf{g}_0 (\lambda)$ is an arbitrary $n$-dimensional, $\lambda$-dependent vector and we have introduced the following matrix polynomials: 
\begin{equation}
\mathbf{L}_i (\lambda) = \mathbf{A}_{1,i}\lambda + \mathbf{A}_{0,i}\, . 
\end{equation}
The reduction to Fuchsian form performed before was necessary in order to have a regular matrix polynomial $\mathbf{L}_0 (\lambda)$, such that $\det(\mathbf{L}_0 (\lambda)) \neq 0$. To find the logarithm-free series solution, we make the ansatz 
\begin{equation}
\mathbf{\widetilde{I}} (x,\lambda,\mathbf{g}_0) = x^{\lambda} \sum_{i\geq 0}\mathbf{g}_i (\lambda) x^i\,,
\end{equation}
where $\mathbf{g}_i(\lambda)$ are arbitrary $n$-dimensional, $\lambda$-dependent vectors, and substitute it into Eq.~\eqref{nonhom-resonance-free}. Extracting coefficients at equal powers of $x$ on both sides of the resulting equation, we obtain
\begin{equation}
\mathbf{L}_0 (\lambda + i)\mathbf{g}_i (\lambda) = -\sum_{j=1}^i\mathbf{L}_j (\lambda + i -j)\mathbf{g}_{i-j}(\lambda)\, , \label{recursion-resonance-free}
\end{equation} 
for $i\geq 1$.
If $\lambda = \lambda_1$ and $\mathbf{g}_0 (\lambda_1)$ is such that $\det(\mathbf{L}_0 (\lambda_1)) = 0$, $\mathbf{L}_0 (\lambda_1)\mathbf{g}_0 (\lambda_1) = \mathbf{0}$, and $\det (\mathbf{L}_0 (\lambda_1 + i))\neq 0$ for $i\geq 1$, then the found solution is also the solution of our initial homogeneous equation in Eq.~\eqref{BarkatouSystem}. The last condition, $\det (\mathbf{L}_0 (\lambda_1 + i))\neq 0$, implies that $\lambda_1$ is a non-resonant eigenvalue of matrix $\mathbf{L}_0 (\lambda)$, and we can solve Eq.~\eqref{recursion-resonance-free} for $\mathbf{g}_i (\lambda_1)$. For our system, we have seven different eigenvalues\footnote{By eigenvalues, we mean solutions of the equation $\det (\mathbf{L}_0 (\lambda)) = 0$.} of matrix $\mathbf{L}_0 (\lambda)$,
\begin{equation}
\lambda \in \left\{-1+\ep , -\frac{1}{2} + \ep , \ep , \ep + \frac{1}{2}, -2 + 2\ep, -2 + 3\ep , -1 + 3\ep \right\}\,,  \label{lambdaEigenvalues}
\end{equation}
with algebraic multiplicities $m_a(\lambda)$,
\begin{eqnarray}
  &&m_a (-1+\ep) = 3\,,\qquad m_a \left(-\frac{1}{2} + \ep\right) = 2\,,\qquad
  m_a (\ep) = 1\,,\qquad m_a \left(\frac{1}{2}+\ep\right) = 1\,,
\nonumber\\
  &&m_a (-2 +2\ep) = 1\,,\qquad m_a (-2+3\ep) = 1\,,\qquad m_a (-1+3\ep) = 1\,. 
\end{eqnarray}
The algebraic multiplicity of eigenvalue $\lambda$ is the number of times it shows up in the polynomial $\det(\mathbf{L}_0 (\lambda))$. Another characteristics associated with eigenvalue $\lambda$ is its geometric multiplicity $m_g(\lambda)$. The latter is defined as the dimension of the eigenspace associated with $\lambda$ such that $\mathbf{L}_0 (\lambda)\mathbf{v} =\mathbf{0}$ with $n$-dimensional eigenvector $\mathbf{v}$. In our case, the geometric multiplicities are equal to the algebraic ones, which implies that we have as many Jordan blocks in matrix $\mathbf{L}_0(\lambda)$ with zero eigenvalue as there are eigenvectors for $\lambda$, and each has size 1. Also, this implies that, for non-resonant eigenvalues, we have as many logarithm-free solutions as corresponds to the number of their geometric multiplicities. In general, one should consider root polynomials associated with the eigenvalues $\lambda$ for each partial multiplicity and account for solutions containing logarithms (see Ref.~\cite{BarkatouFrobenius} for more details).

From the expressions for the eigenvalues $\lambda$ in Eq.~\eqref{lambdaEigenvalues}, we see that we have resonant eigenvalues $\lambda = -1+\ep, -1/2+\ep, -2+3\ep$, and the solution presented above is not valid in these cases. In principle, we could use balance transformations \cite{RomanBalances} to get rid of the resonant eigenvalues in our system from the very beginning. Still, the method of Ref.~\cite{BarkatouFrobenius} allows us to find the corresponding series solutions even in the case of resonances. Indeed, following Ref.~\cite{BarkatouFrobenius}, let us consider the resonant eigenvalue $\lambda_1$ such that $\operatorname{Re}\lambda_1 < \cdots < \operatorname{Re}\lambda_r$ and $\lambda_i - \lambda_1\in \mathbb{N}^*$. Setting $\alpha =\alpha (\lambda_1) = \sum_{i=2}^r m_a (\lambda_i)$,\footnote{In our case, all values of $\alpha$ for resonant eigenvalues are equal to unity.}
let us consider the modified non-homogeneous system,
\begin{equation}
\mathcal{L}(\mathbf{\widetilde{I}}) = (\lambda - \lambda_1)^{\alpha} \mathbf{L}_0 (\lambda) \mathbf{g}_0 (\lambda) x^{\lambda}\,, \label{nonhom-resonance}
\end{equation}
where $\mathbf{g}_0 (\lambda)$ is again an arbitrary $n$-dimensional, $\lambda$-dependent vector. Substituting the ansatz 
\begin{equation}
\mathbf{\widetilde{I}} (\lambda, x, \mathbf{g}_0) = \left[
(\lambda - \lambda_1)^{\alpha}\mathbf{g}_0(\lambda) + \sum_{k\geq 1}\mathbf{g}_k (\lambda) x^k\right] x^{\lambda}\,,
\end{equation}
into Eq.~\eqref{nonhom-resonance}, we obtain the following recurrence relation for the coefficients of the logarithm-free series solution:
\begin{equation}
\mathbf{L}_0 (\lambda + k)\mathbf{g}_k (\lambda) = -\sum_{j=1}^{k-1}\mathbf{L}_j (\lambda + k -j)\mathbf{g}_{k-j}(\lambda) - (\lambda - \lambda_1)^{\alpha}\mathbf{L}_k (\lambda)\mathbf{g}_0 (\lambda)\,, \label{recurrence-resonace}
\end{equation}
for  $k\geq 1$.
The term proportional to $(\lambda-\lambda_1)^{\alpha}$ on the right-hand side of
Eq.~\eqref{recurrence-resonace} guarantees that we can always solve Eq.~\eqref{recurrence-resonace} for $\mathbf{g}_k(\lambda)$ in the vicinity of $\lambda = \lambda_1$ and that the obtained solution is regular at $\lambda=\lambda_1$. The found solution is, however, a multiple of a regular solution associated with the non-resonant eigenvalue $\lambda_r$ calculated before. To find linearly independent solutions, we first find $m_g(\lambda_1)$ independent eigenvectors $\mathbf{g}_{0,i}(\lambda_1)$ of the system $\mathbf{L}_0(\lambda_1)\mathbf{g}_{0,i}(\lambda_1) =\mathbf{0}$. Then, the sought solutions are given by the derivatives
$d^{\alpha}\mathbf{\widetilde{I}} (\lambda, x,\mathbf{g}_{0,i})/d\lambda^{\alpha}$.\footnote{This is only valid in the case $m_a(\lambda_1) = m_g (\lambda_1)$. See Ref.~\cite{BarkatouFrobenius} for a discussion of the general situation.}

In this way, we obtain $n=10$ linearly independent solutions of the system
in Eq.~\eqref{BarkatouSystem} altogether. To fix ten constants in the general solution, we account for boundary conditions at $x=0$. The latter are obtained
using asymptotic expansions of the Feynman diagrams at $x = 0$, properly taking
subgraphs into account, and read 
\begin{eqnarray}
  e^{2\ep\gamma_{E}}\left(M^{2}\right)^{2\ep-2}I_{0000011}&= & -\frac{1}{\ep^{2}}-\frac{2}{\ep}-3-\zeta_{2}+\OO(\ep)\,,
  \nonumber\\
  e^{2\ep\gamma_{E}}\left(M^{2}\right)^{2\ep-2}I_{0000101}&= & -\frac{x}{\ep^{2}}+\frac{x(\ln x-2)}{\ep}+x\left(2\ln x-\frac{\ln^{2}x}{2}-3-\zeta_{2}\right)+\OO(\ep)\,,
  \nonumber\\
  e^{2\ep\gamma_{E}}\left(M^{2}\right)^{2\ep-1}I_{0000111}&= & -\frac{1}{\ep^{2}}-\frac{3}{\ep}-7-\zeta_{2}+\OO(\ep)\,,
  \nonumber\\
  e^{2\ep\gamma_{E}}\left(M^{2}\right)^{2\ep}I_{0000121}&= & -\frac{1}{2\ep^{2}}-\frac{1}{2\ep}-\frac{1}{2}-\frac{\zeta_{2}}{2}+\OO(\ep)\,,
  \nonumber\\
  e^{2\ep\gamma_{E}}\left(M^{2}\right)^{2\ep}I_{0000211}&= & -\frac{1}{2\ep^{2}}+\frac{1}{\ep}\left(\ln x-\frac{1}{2}\right)+\frac{3}{2}-\frac{\zeta_{2}}{2}-\frac{\ln^{2}x}{2}+\OO(\ep)\,,
  \nonumber\\
  e^{2\ep\gamma_{E}}\left(M^{2}\right)^{2\ep-1}I_{0001011}&= & -\frac{1}{\ep^{2}}-\frac{3}{\ep}-7-\zeta_{2}+\OO(\ep)\,,
  \nonumber\\
  e^{2\ep\gamma_{E}}\left(M^{2}\right)^{2\ep}I_{0001111}&= & -\frac{1}{2\ep^{2}}+\frac{1}{\ep}\left(\ln x-\frac{1}{2}\right)+\frac{3}{2}-\frac{\zeta_{2}}{2}-\frac{\ln^{2}x}{2}+\OO(\ep)\,,
  \nonumber\\
  e^{2\ep\gamma_{E}}\left(M^{2}\right)^{2\ep}I_{0002011}&= & -\frac{1}{2\ep^{2}}+\frac{1}{\ep}\left(\ln x-\frac{1}{2}\right)+\frac{3}{2}-\frac{\zeta_{2}}{2}-\frac{\ln^{2}x}{2}+\OO(\ep)\,,
  \nonumber\\
  e^{2\ep\gamma_{E}}\left(M^{2}\right)^{2\ep}I_{0010111}&= & -\frac{1}{2\ep^{2}}+\frac{1}{\ep}\left(\ln x-\frac{1}{2}\right)+\frac{3}{2}-\frac{\zeta_{2}}{2}-\frac{\ln^{2}x}{2}+\OO(\ep)\,,
  \nonumber\\
e^{2\ep\gamma_{E}}\left(M^{2}\right)^{2\ep}I_{0011101}&= & \frac{1}{2\ep x}+\frac{1}{2x}(1-\ln x)+\OO(\ep)\,.
\end{eqnarray} 
Having found the Frobenius solution for the vector of master integrals in Eq.~\eqref{mastersvector}, we may use the known basis of occurring sums to reconstruct the series solutions presented in Subsection~\ref{seriesRep}.

\end{document}